\documentclass[fleqn,usenatbib]{mnras}

\usepackage{newtxtext,newtxmath}

\usepackage[T1]{fontenc}

\DeclareRobustCommand{\VAN}[3]{#2}
\let\VANthebibliography\thebibliography
\def\thebibliography{\DeclareRobustCommand{\VAN}[3]{##3}\VANthebibliography}

\usepackage{graphicx}	
\usepackage{amsmath}	
\usepackage{enumitem}

\newcounter{g}
\renewcommand{\theg}{G\arabic{g}}

\newcommand{\g}[1]{%
  \refstepcounter{g}%
  \label{eq.#1}%
  \theg%
}

\title[Unresolved binaries in the VVV b294 tile]{Identification of unresolved binaries in the VVV b294 tile: A Pro-Am collaboration using the Virtual Observatory}

\author[Cort\'es-Contreras et al.]{
M. Cort\'es-Contreras,$^{1}$\thanks{E-mail: mcortescontreras@ucm.es}
P. Cruz,$^{2}$
E. Solano,$^{2}$
J. L. Navarro,$^{3}$
T. Poyato Romero,$^{4,5}$
R. Benavides Palencia,$^{4}$
\newauthor{
C. Cagigal-Olay,$^{6}$
M. Zamora Rodríguez,$^{7}$
A. Ginard,$^{5,8}$
S. Langa,$^{9}$
J. González-Edo,$^{10}$
P. L'Huillier Seguel,$^{5}$
}
\newauthor{
E. M. Nicola,$^{8}$
V. Pallares-López,$^{4}$
M. A. Almela Rubert,$^{10}$
J. A. Cano Díez,$^{11}$
J. E. Donate Lucas,$^{12}$
}
\newauthor{
N. Echevarria Sainz-Ezquerra,$^{13}$
A. Elbaz-Sanz,$^{14}$
J. M. Espinosa Gallardo,$^{15}$
G. Gurrea-Ysasi,$^{16}$
}
\newauthor{
J. C. Mora,$^{17,18}$
C. Morales-Socorro,$^{19}$
J. M. Perales,$^{8}$
A. Romar$^{9}$
}
\\
$^{1}$Departamento de F\'isica de la Tierra y Astrof\'isica \& IPARCOS-UCM (Instituto de F\'isica de Part\'iculas y del Cosmos de la UCM), Facultad de Ciencias F\'isicas, \\ Universidad Complutense de Madrid, 28040 Madrid, Spain \\
$^{2}$ Centro de Astrobiolog\'ia (CSIC-INTA), Camino Bajo del Castillo s/n, ESAC Campus, 28692 Villanueva de la Ca\~nada, Spain\\
$^{3}$Asociación Astronómica AstroHenares, Centro de Recursos Asociativos El Cerro, C/ Manuel Azaña S/N, 28823 Coslada, Spain\\
$^{4}$Agrupación Astronómica de Córdoba, C/ Arquitecto Sáenz de Santamaría 14, 14010 Córdoba, Spain\\
$^{5}$Agrupación Astronómica de Sabadell, C/ Prat de la Riba 116, 08206 Sabadell, Spain\\
$^{6}$Agrupación Astronómica de Madrid, C. Albendiego 22, 28029 Madrid, Spain\\
$^{7}$Agrupación Astronómica Rías Baixas, Pontevedra, Spain
$^{8}$Asociación Astronómica de Mallorca, C/ Sa Lluna 8, 07314 Caimari, Mallorca, Spain\\
$^{9}$Independent researcher, España\\
$^{10}$Societat Astronòmica de Castelló, Antiguo Cuartel Tetuán 14,  Edificio Asociaciones 2º, 12004 Castelló, Spain\\
$^{11}$Asociación Astronómica Cielos Despejados, Oviedo, Spain\\
$^{12}$Agrupación Astronómica de Cuenca, Plaza de la Merced 1,  16001 Cuenca, Spain\\
$^{13}$Asociación Astronómica de Enkarterri Betelgeuse, Av. Lanzagorta 40 bajo, 48860 Zalla, Bizkaia, Spain\\
$^{14}$Instituto de Ciencias del Espacio (ICE-CSIC), Carrer de Can Magrans, 08193 Cerdanyola del Vallès, Barcelona, Spain\\
$^{15}$Asociación Astronómica de Tomares (ASTROMARES), Avda. de Andalucía S/N., 41940 Tomares, Spain\\
$^{16}$Instituto COMAV, Ciudad Politécnica de la Innovación, Universitat Politècnica de València, Camino de Vera S/N, 46022, Valencia, Spain\\
$^{17}$Inter-University Centre for Astronomy and Astrophysics (IUCAA), Post Bag 4 , Ganeshkhind, Savitribai Phule Pune University Campus, Pune 411 007, India\\
$^{18}$Universidad de Chile, Av. Libertador Bernardo O'Higgins 1058, Santiago de Chile\\
$^{19}$Asociación Astronómica y Educativa de Canarias "Henrietta Swan Leavitt", C/Artemi Semidán 3, 35460 Las Palmas, Spain\\
}

\date{Accepted 2026 July 27. Received --; in original form --}

\pubyear{\the\year{}}

\begin{document}
\label{firstpage}
\pagerange{\pageref{firstpage}--\pageref{lastpage}}
\maketitle

\begin{abstract}
{Stellar binaries and multiples are common objects. Their identification and characterization are primarily limited by the angular resolution of current instrumentation, survey depth and dynamic range. Towards the Galactic plane, stellar crowding  further complicates these studies.}
{We aim to identify unresolved binaries with good astrometric solutions and M dwarf primaries in the b294 field from the Vista Variables in the V\'ia L\'actea (VVV) survey using publicly available data.
We examine the catalogue of M dwarfs within 500\,pc in the b294 tile, employing binarity tracers based on statistical parameters from \textit{Gaia} Data Release 3, multi-band photometry, and information in the literature. We reassess the astrometric and photometric parameters typically used for binary detection, establishing new boundaries and validating existing ones in the dense stellar field examined.
We compare these results with those obtained in the solar vicinity. 
We present 990 unresolved binary candidates and 43 previously known binaries, including five possible triple systems. These represent the 13\% of the catalogue, rising to 14.9\% in well-covered regions. Binary-model SED fitting yields effective temperatures for 98 primary stars, with the properties of their secondary candidates indicating that nearly half may be substellar objects.} We select 120 of the most promising binary candidates. Tangential velocities, where calculated, place 95\% of the studied binary candidates within the young disc population ($\tau \lesssim1$\,Ga). We demonstrate that the influence of nearby \textit{Gaia} detections on the statistical parameters is small, allowing constraints similar to those that are applied in less crowded regions.
\end{abstract}

\begin{keywords}
surveys -- virtual observatory tools -- astrometry -- stars: binaries: general -- stars: binaries: close -- stars: low-mass
\end{keywords}



\section{Introduction}\label{sec:introduction}

The search for binary stars is an almost ancient topic of research. Since the very first binaries identified centuries ago, hundreds of them have been detected and characterized at a large range of masses and separations (e.g., \citealt{DM91, Raghavan10, FM92, Payo23}). Stellar multiplicity plays a crucial role in stellar evolution.
Theoretical models and observations strongly support that most stars form in multiple systems \citep{Bate12, Tokovinin14, Offner23}, although continuous small and dissipative encounters may disrupt them with time \citep{RK82,Borkovits22,Yip23}. Dynamical evolution can also favor the formation of hierarchical systems with more than two stars with a number of stable configurations \citep{Evans68, Tokovinin14,BT24}. The environment summed to the age strongly affect such configurations. In the solar neighborhood the vast majority of systems found are stable binaries, with multiplicity fractions that decrease with the stellar mass of the primary \citep{Payo23}.

Stellar regions towards the Galactic plane present an heterogeneous environment with high stellar density and the presence of dust that obscure stellar emission. In this context, the VVV survey \citep{Minniti10,Minniti18} provides multi-epoch IR photometry in the bulge and Galactic plane with the purpose of mapping this region. Focusing on identifying M dwarfs, \cite{PCruz23} (hereafter PC) performed an astrometric and photometric procedure to identify and characterize M dwarfs in the b294 tile in the inner bulge region with the Virtual Observatory (VO). The VO is a framework that integrates astronomical data archives and software tools and services, providing support for astronomical research.

This particular tile was selected to complement the search of M dwarfs using VVV data carried out by \cite{RojasAyala14} in a border region of the bulge, where the star density is lower. In contrast, the b294 tile is located in a denser region of the bulge (it contains near 2.2$\cdot$10$^6$ objects per square degree detected in the near-infrared bands of VVV, an order of magnitude larger than the region studied by \citealt{RojasAyala14}) and spans the following coordinates: $3.1^\circ < l < 4.7^\circ$ and $-3.8^\circ < b < -2.5^\circ$.
Compared with the optical output of \textit{Gaia} DR3 \citep{Gaia16, GaiaeDR3} in the region of study, the PC work significantly increased the number of known M dwarfs. The catalogue is restricted by design to stars with {\tt RUWE} < 1.4 to ensure reliable astrometric solutions. This criterion naturally excludes sources whose astrometry may be perturbed by a physical companion. Still, a sample of potential unresolved binary candidates was identified.
The present work aims to assess the PC catalogue in terms of unresolved multiplicity that survives the {\tt RUWE} criterion. Our goal is to extend the study of such systems in this mass range and region of the sky.
The M-dwarf sample from PC provides an opportunity to test a methodology based entirely on publicly available data, intended to be applied for further multiplicity research in crowded fields. This analysis will allow us to evaluate its strengths and limitations, while also identifying the most promising targets for follow-up observations. We additionally aim to provide a lower limit  binary fraction of M dwarfs up to 500\,pc toward the Galactic plane, enabling a direct comparison with field populations. A secondary objective is to explore the capabilities of current astronomical surveys in crowded environments.

This work is the result of join efforts between professional and amateur astronomical communities, and is structured as follows. In Section~\ref{sec:sample}, we introduce the sample of work. In Sections~\ref{sec:methodology} and \ref{sec:identificationbinaries}, we describe the methodology applied and the binary candidates obtained after that. The analysis and results are presented in Section~\ref{sec:analysisandresults}, and the summary of this work is contained in Section~\ref{sec:summary}.

\section{Sample}
\label{sec:sample}

We use as input the sample of 7\,925 M dwarfs identified in \cite{PCruz23} in the b294 tile of the VVV survey (\cite{Minniti10,Minniti18}), hereafter the PC catalogue. The sample includes stars located within 500\,pc from the Sun and is constructed from parallaxes, or proper motions and colour cuts based on the colour ranges presented in \cite{RojasAyala14}. 

\begin{figure}
\centering
\includegraphics[width=0.45\textwidth]{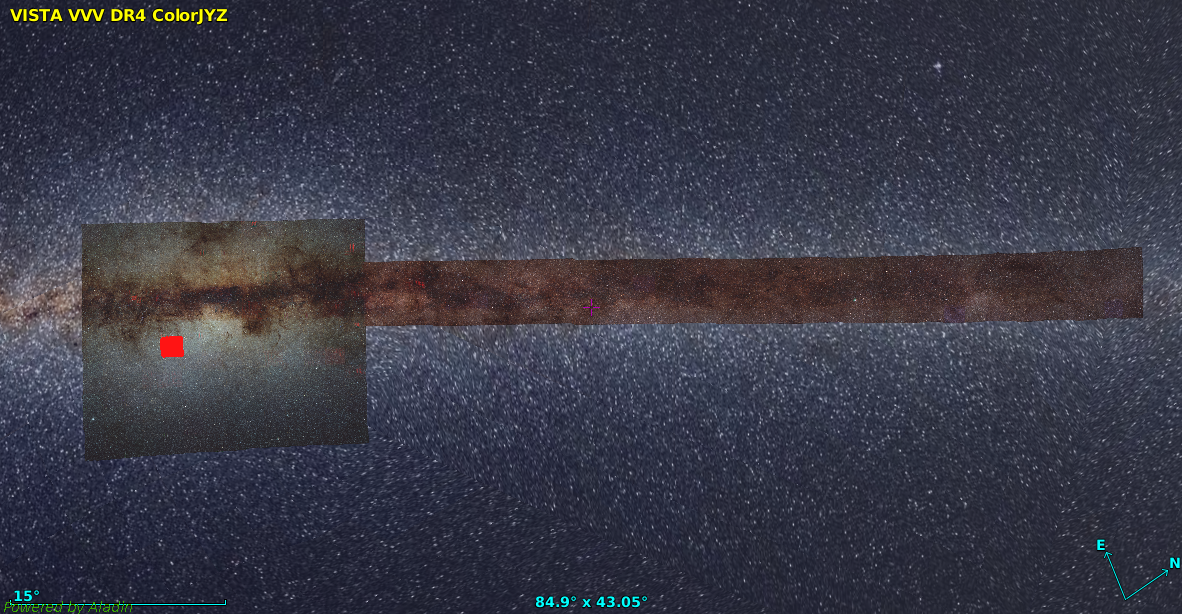}
\caption{The sky distribution of the sample of 7\,925 M dwarfs in the b294 tile is shown in red. The shadowed area corresponds to the VVV coverage.}
\label{fig.sky}
\end{figure}

\begin{figure}
\centering
\includegraphics[width=0.45\textwidth]{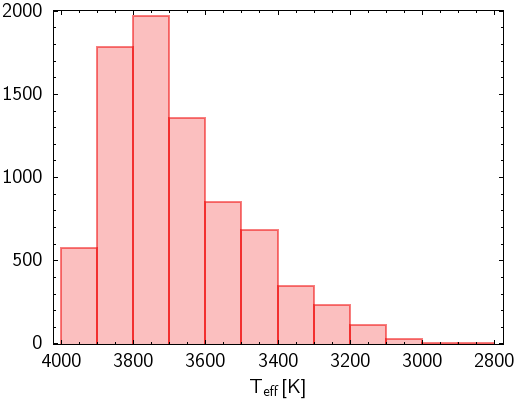}
\caption{Distribution of effective temperatures obtained for the input sample. The used grid of models has a step in temperature of 100\,K, which was adopted here as the bin size.}
\label{fig.histteff}
\end{figure}

Its sky position with respect the Galactic plane is shown in Figure~\ref{fig.sky}. The spectral types range from M0\,V to $\sim$M6\,V according to their effective temperatures, which were obtained using VOSA\footnote{\url{http://svo2.cab.inta-csic.es/theory/vosa/}} (Virtual Observatory SED Analyzer, \citealt{Bayo08}) and the BT-Settl CIFIST models \citep{Allard11}. The equivalence effective temperatures -- spectral types is based on Table~6 in \cite{Cif20} and is shown in Figure~\ref{fig.histteff}.

The PC catalogue is accessible from the SVO archive\footnote{\url{http://svocats.cab.inta-csic.es/mdwarfs_vvv/}}.
Coordinates and $ZYJHKs$ magnitudes are taken from VVV, while stellar parameters (T$_{\rm eff}$, L$_{bol}$, and radii) are those derived with VOSA in \cite{PCruz23}. Distances are either from {\sl Gaia} DR3 or spectophotometric derived by the authors.

\section{Methodology}
\label{sec:methodology}

The identification of unresolved binary candidates is addressed using a combination of complementary criteria, as none of them alone can be considered as a direct and irrefutable evidence of the presence of a binary companion. Therefore, we evaluate a series of indicators in order to identify the most promising unresolved binary candidates in our sample. 

\subsection{IR excess}\label{sec.IRexcess}

Aiming at identifying close unresolved binaries, we first pay attention to the {\tt flag\_IR} column included in the PC catalogue, which is set to TRUE for the 1\,228 stars that present infrared excess in their SEDs explored with VOSA. As stated by the authors, this excess could be related to a wrong photometric counterpart used to build the SED, contaminated photometry of our target due to a close visual neighbor or an unresolved bound companion. 
The crowding of the field prevents us from reliably confirming the case attending uniquely to the separation between the photometric counterpart and the input coordinates.

For addressing this issue, we started up a Pro-Am collaboration (i.e., a collaboration between professional and amateur astronomers), described in Appendix~\ref{app.proam}. It is based on visual inspection of the field with the Aladin Sky Atlas \citep{Bonnarel00} carried out several times to reach overall agreement on the identification of stellar contaminants. We used the VVV DR4 colour image of the field, the VVV coordinates of the targets and the position of sources in the GLIMPSE Source Catalog \citep{Glimpse}. This catalogue provides photometry in the I1 to I4 bands of the Infrared Array Camera (IRAC) onboard the Spitzer Space Telescope  \citep{IRAC} centered at approximately 3.6, 4.5, 5.8 and 8.0\,$\mu$m at which VOSA detected the IR excesses.

This visual inspection allows us to directly discard those targets contaminated or incorrectly matched (flagged ``C'') and to keep those with a correct IRAC counterpart to be investigated further (flagged ``Y''). Sources without sufficient consensus are flagged as ``I'' (inconclusive).

\subsection{Gaia DR3 indicators} \label{sec.gaiaindicators}

To complement the previous analysis, we revised our use of \textit{Gaia} DR3 statistical parameters known to be valuable for assessing the reliability of the astrometric data. The astrometric solution could be affected by the astrometric or photometric disruption of a close unresolved companion. \citet{Fabricius21} and \citet{Cifuentes25} proved that certain parameters from \textit{Gaia} provide a complementary approach for identifying potential close binary systems. All of them are properly described in the Gaia DR3 documentation\footnote{\url{https://gea.esac.esa.int/archive/documentation/GDR3/index.html}} but are shortly described here:

\begin{itemize}

    \item {\tt RUWE} is the Renormalised Unit Weight Error, an indicator of the quality of the astrometric fit. Values close to 1 indicate a good single-star solution, while values greater than 1.4 often suggest binarity or other issues in the astrometric model.

    \item {\tt ipd\_gof\_harmonic\_amplitude} is the amplitude of the first harmonic of the image parameter determination (IPD) goodness-of-fit variation. Large values may indicate non-single or extended sources.

    \item {\tt ipd\_frac\_multi\_peak} is the fraction of transits with multiple peaks detected in the IPD process, often pointing to blended or close multiple systems.

    \item {\tt rv\_chisq\_pvalue}: p-value of the chi-squared test for the radial velocity model fit. Low values indicate a poor fit to the data.

    \item {\tt rv\_renormalised\_gof} is the renormalised goodness-of-fit statistic for the radial velocity solution, with large values suggesting inconsistency between the model and the observed radial velocities.

    \item {\tt rv\_nb\_transits} is the number of RV transits used to derive the median radial velocity.

    \item {\tt radial\_velocity\_error} is the formal uncertainty of the median radial velocity.

    \item {\tt duplicated\_source} corresponds to a boolean flag indicating whether the source appears more than once in the Gaia processing, which may hint at ambiguous detections.

    \item {\tt non\_single\_star} flags sources modeled as non-single (e.g. astrometric, spectroscopic, or eclipsing binaries) in Gaia’s non\_single\_star catalogue.

\end{itemize}.

The criteria provided in Table~3 for the detection of unresolved sources by \cite{Cifuentes25} are summarized in Table~\ref{tab.Gaiacriteria}.

\begin{table}
 \centering
 \caption{{Astrometric and photometric indicators from {\it Gaia} DR3.}}
 \small
 \begin{tabular}{ll}
 \hline 
 \hline
 \noalign{\smallskip}
Criterion ID  & Criterion \\

 \noalign{\smallskip}
 \hline
 \noalign{\smallskip} 

\g{ruwe}    &   {\tt RUWE} > 2  \\
 \noalign{\smallskip}
 \g{gof}    &   {\tt ipd\_gof\_harmonic\_amplitude} > 0.1 \& {\tt RUWE} > 1.4    \\
 \noalign{\smallskip}
 \g{multipeak}  & {\tt ipd\_frac\_multi\_peak} > 30 \\
 \noalign{\smallskip}
 \g{chisq}  &   {\tt rv\_chisq\_pvalue} < 0.01 \& \\
                    &   {\tt rv\_renormalised\_gof} > 4 \& \\
                    & {\tt rv\_nb\_transits} $\geq$ 10 \\
 \noalign{\smallskip}
 \g{rv}  & {\tt radial\_velocity\_error} $\geq$ 10\,{\rm km\,s$^{-1}$} \\
 \noalign{\smallskip}
  \g{dupsource}  & {\tt duplicated\_source} = 1 \\ 
 \noalign{\smallskip}
\g{nonsingle}   &    {\tt non\_single\_star} $\neq$ 0 \\
 \noalign{\smallskip}
 \g{epsi}               & {\tt ipd\_gof\_harmonic\_amplitude} > 0.1 \& \\
                & {\tt astrometric\_excess\_noise} > 1.4 \& \\
                & {\tt astrometric\_excess\_noise\_sig} > 2 \\
\noalign{\smallskip}
 \hline
 \end{tabular}
\label{tab.Gaiacriteria}
\end{table}

Due to the construction of the PC catalogue, no star with {\tt RUWE} over 1.4 was kept since only stars with good astrometric solutions were selected. Hence, criteria~\ref{eq.ruwe} and \ref{eq.gof} would not provide any candidate here. Alternatively, there are other parameters to be considered that are often not used because of redundancy with {\tt RUWE}: the {\tt astrometric\_excess\_noise} ($\epsilon$) and the {\tt astrometric\_excess\_noise\_sig}. The former represents an additional noise term added to the astrometric model to account for residuals not explained by the standard single-star five-parameter solution. Large values indicate that the source does not fit well a single-star astrometric model, possibly due to binarity, extended structure, or other issues. The latter stands for the significance of the astrometric excess noise, defined as a signal-to-noise ratio. As a reference, typical values of the astrometric excess noise for stars ranging between $G$-band magnitudes 16 and 20\,mag (magnitude interval for the M dwarfs in our sample), range from 0.285\,mas to 0.976\,mas at 19\,mag and then rises to 1.801\,mas at 20\,mag (Table~4 in \citealt{Lindegren21}). According to these authors, this parameter is considered statistically significant when {\tt astrometric\_excess\_noise\_sig} is larger than 2. Nonetheless, they state that noise is expected to increase in dense stellar regions due to background sources, and typical values need to be revisited in this work. 

In the following, we adopt the criterion \ref{eq.epsi} in Table~\ref{tab.Gaiacriteria} as a substitute for  \ref{eq.ruwe} and \ref{eq.gof}, where the limit of 1.4 in the {\tt astrometric\_excess\_noise} is defined as the average plus the standard deviation of the median values provided in Table~4 in \cite{Lindegren21}. As a reference, this number is consistent with -- and more conservative than -- the upper end within 1$\sigma$ of the median value of $0.44^{+0.64}_{-0.26}$\,mas (1.08\,mas) reported for a sample of X-rays close binaries in \cite{Gandhi22}. It also excludes low noise values (below 1\,mas), which may correspond to partially resolved double stars, but can also reflect contamination from source variability and systematic effects, as noted by the authors.

At least one of these six (\ref{eq.multipeak} to \ref{eq.epsi}) criteria must be satisfied for a star to be considered a potential binary candidate, pending further confirmation.

\subsection{Photometric variability}\label{sec.photvar}

Photometric variability can arise from several physical mechanisms, including eclipses with objects of stellar, substellar or planetary nature, where periodic flux variations are produced by mutual occultations; ellipsoidal variability caused by tidal distortion of stars in close binary systems; planet-like transit signals; and intrinsic stellar activity, such as rotational modulation by star spots and flares. Consequently, photometric variability may indicate the presence of a close companion that provokes periodic fluctuations in the measured fluxes. However, the origin of the variability must be established through the analysis of time-series observations with sufficient temporal baseline and cadence. Both the VVV and \textit{Gaia} surveys provide light curves for a subset of the sample, which are used as additional diagnostic for assessing binarity alongside the \textit{Gaia}-based criteria listed in Table~\ref{tab.Gaiacriteria}.

On the one hand, \cite{PCruz23} searched for photometric variability in 4\,951 M dwarfs with VVV light curves out of the total sample of 7,925 stars. They evaluated the reduced chi-square indicator as in \cite{Botan21} to define two subsets of M dwarfs with periodic signals: $\chi^2>2$ and $\chi^2 \sim 1$ indicate variability more likely associated with the presence of a stellar companion and variability consistent with a planet-like transiting object or intrinsic stellar variability, respectively.

On the other hand, the \textit{Gaia} observing strategy permits a photometric variability analysis for million of stars in the $G$, $RP$, and $BP$ bands. Details of the variable source processing and analysis can be found in \cite{Eyer23}. The results of such analysis are summarized in the {\tt phot\_variable\_flag}, set to "VARIABLE" when the star shows photometric variability.
Although we use variability as a binarity indicator, further analysis should be carried out in order to clarify its nature.

\subsection{Literature}

We recovered the multiplicity information available in the literature by first crossmatching our catalogue with the SIMBAD astronomical database \citep{Wenger00} and checking out the \textit{Object type} column. This column presents a hierarchical classification, which describes the physical nature of the object. We focus on labels referring to binarity, such as EB${^*}$, SB${^*}$ or XB${^*}$ (eclipsing, spectroscopic or X-ray binary, respectively).

Aiming at being as comprehensive as possible, we independently crossmatched our sample with the Washington Double Star catalogue (WDS, \citealt{Mason01}) and with several catalogues from the OGLE project \citep{Paczynski86,Udalski93}, in order to identify known systems that may not be fully represented in the SIMBAD database. The OGLE project, originally designed to search for dark matter through microlensing events, has also been prolific in identifying contact binaries and variable stars.

We also performed an independent crossmatch with the \textit{Non-single stars} tables from \textit{Gaia} \citep{Arenou23}, which include sources exhibiting significant deviations from a single-star model such as unresolved astrometric, spectroscopic, and eclipsing binaries.

\section{Identification of binary candidates}
\label{sec:identificationbinaries}

The number of stars flagged as positive in each of the following subsections are summarized in Table~\ref{tab.summarynumbers}.

\begin{table*}
 \centering
 \caption{Number of stars that satisfy each astrometric or photometric criterion. }
 \begin{tabular}{cccccccccccccc|c}
 \hline 
 \hline
 \noalign{\smallskip}
   IR &   G3 & G4 & G5 & G6 & G7   &   G8  &  $\chi^2$  & GVar  &   GBin     &   Ovar &   OBin &  SVar &   SBin   & Total   \\

 \noalign{\smallskip}
 \hline
 \noalign{\smallskip} 
   98  &   520    &   0   &   0   &   12  &   0   & 431 & 47 &  7  & 37   &  0 &  6    &  6 &   17    &   1033$^a$\\
 \noalign{\smallskip} 
\noalign{\smallskip}
 \hline
 \end{tabular}
\vspace{2mm}
\begin{minipage}{\textwidth}
 {\footnotesize
 $^{a}$ Total number of stars after accounting for stars with multiple positive flags.
 }
 \end{minipage}
 \label{tab.summarynumbers}
\end{table*}

\subsection{IR excess}\label{subsecIR}

Out of the 1228 stars with IR excess, we classify 98 as stars with a confirmed excess in their SEDs (``Y'') and 977 as contaminated (``C''), all of them with an agreement over 60\% in the visual inspection as described in Sect.~\ref{sec.IRexcess}. The remaining 153 stars are flagged as inconclusive (``I'') due to strong disagreement among the five to six independent classifications, primarily reflecting the difficulty in discerning the source of the excess.
After this review, 8\% of the studied stars remain as binary candidates.

\begin{figure}
\centering
\includegraphics[width=0.45\textwidth]{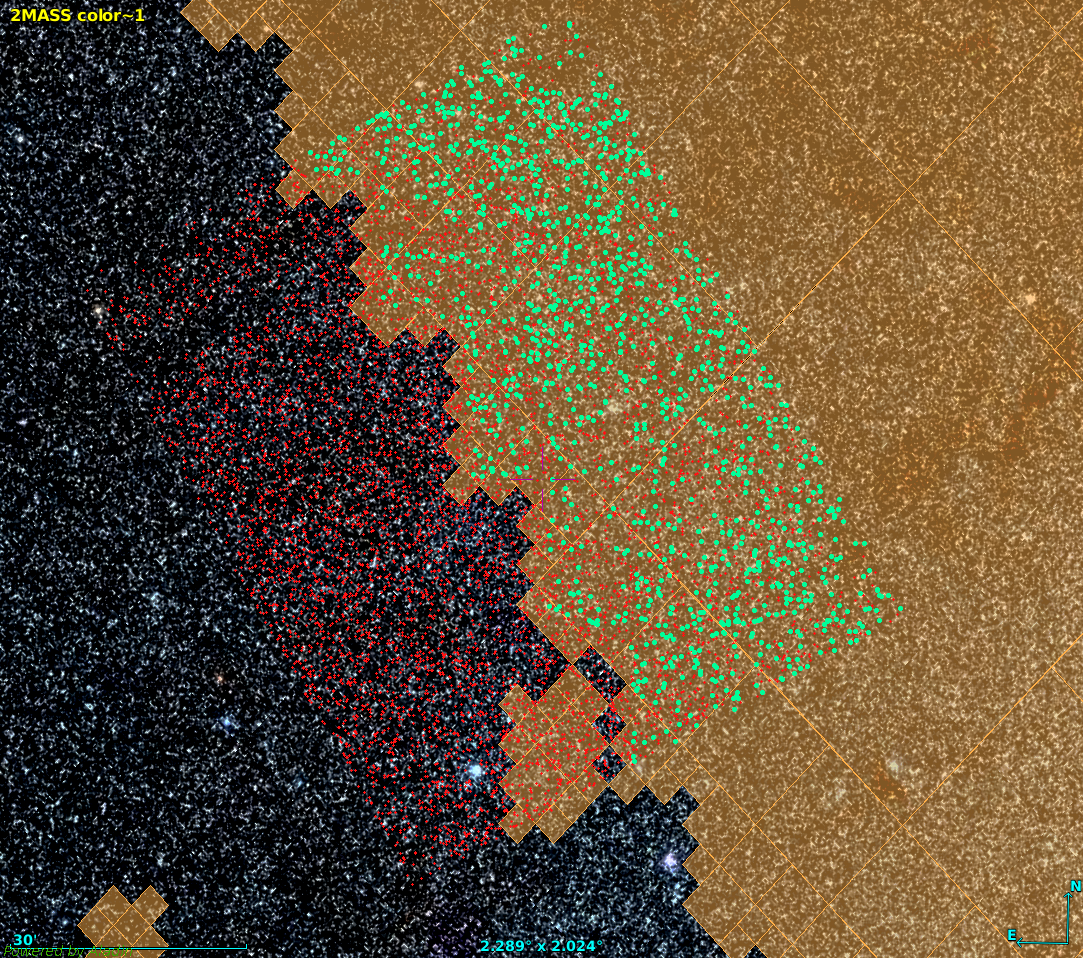}
\caption{Aladin sky view with the 2MASS colour image as a background. Red dots and green circles stand for the M dwarfs in the PC catalogue and the 1228 stars with IR excess, respectively. Orange-shaded region represent the MOC coverage of GLIMPSE.}
\label{fig.mocirac}
\end{figure}

It is worth noting that the 1228 M dwarfs with IR excess in their SEDs are limited to nearly half of the tile analyzed in \cite{PCruz23}, since the GLIMPSE catalogue fails at covering the whole region (see Figure~\ref{fig.mocirac}). 
Assuming symmetry, we would expect to find similar results in the missing half of the field. This will be considered when assessing multiplicity in the PC catalogue.

\subsection{Gaia DR3 indicators}

The Gaia DR3 statistical parameters criteria described in Sect.~\ref{sec.gaiaindicators} were applied to the complete sample of 7925 M dwarfs, regardless of previous results.
Criterion~\ref{eq.multipeak}, ascribed to the fraction of windows for which the algorithm has identified a double peak, is fulfilled by $\sim$7\% of the stars, closely followed by criterion~\ref{eq.epsi}. Criterion~\ref{eq.dupsource}, related to the existence of a duplicated source during data processing, applies to only 12 stars (0.15\%). The rest of the criteria yields no candidates. The number of true cases is expressed for each criterion in  Table~\ref{tab.summarynumbers}.
A total of 853 stars fulfill at least one of these criteria.

\subsection{Photometric variability}

As described above in Sect.~\ref{sec.photvar}, we considered the two subsamples of stars identified in \cite{PCruz23} that show values of $\chi^2 \sim 1$ (20) and $\chi^2 >2$ (27), as they indicate that observed variability could be produced by a stellar or planet-like companion. These 47 objects are grouped together under the ``$\chi^2$'' column in Table~\ref{tab.summarynumbers}.

Concerning the {\sl Gaia} variability flag, we found 44 M dwarfs identified as ``VARIABLE'' out of the 7\,925 M dwarfs.
They are included in \textit{Gaia}'s \textit{Variability} tables, from which we can retrieve more information about them: 25 are classified as ellipsoidal variables with possible compact object secondaries \citep{Gomel23}, eleven as EBs \citep{Mowlavi23}, six as long-period variables \citep{Lebzelter23}, one young stellar object (YSO) \citep{Marton23}, and one RS Canum Venaticorum type variable.
The six long period variables and the YSO are included in Table~\ref{tab.summarynumbers} under ``GVar'', while the remaining 37 stars are included under "GBin".

\subsection{Literature}\label{sec.lit}

None of the M dwarfs in our sample has a counterpart within 1\,arcsec in either the WDS catalogue or the \textit{Gaia}'s \textit{Non-single stars} tables.
We searched for OGLE catalogues overlapping the region of our tile with Aladin using a search radius of 1\,arcsec. We found one contact binary and five non-contact binaries in the catalogue of contact binaries of \citep{Soszynski16}. These six stars are classified in \cite{Bodi21} as detached binaries (one, c $ \leq 0.5$), semi-detached binaries (two, $0.5 < c \leq 0.7$), and ellipsoidal and over contact systems (three, $c>0.7$), as defined by the morphology parameter $c$. 
We include two columns in Table~\ref{tab.summarynumbers}: ``OBin'' and ``Ovar''. Since all OGLE variables are associated to binarity, all of them are included in the ``Obin'' column, ending ``Ovar'' with no stars.

Lastly, we found 44 stars in Simbad within 1\,arcsec. There are 17 stars identified as eclipsing binaries (EB*), one of which is also flagged as variable (V*). We also found two long period variables (LP*) and four additional variables (V*). The remaining 21 stars (44$-$23) have mainly ``Star'' classifications but there are also high proper motion stars (``HighPM*''). Some of them are also identified as near IR source (``NIR'') or ascribed to microlensing events (``Lev'').
We define two columns in Table~\ref{tab.summarynumbers}: ``SBin'' and ``SVar'' which contain confirmed binary stars (EB*) and variables (LP* and V*), respectively. The second sample excludes the one star classified as EB.
We notice that almost all the binaries and variables identified by OGLE and Gaia in the previous exercise are accounted also in SIMBAD.

Based on these findings, the presence of close binary systems in the sample is so far constrained to very close unresolved binaries.

\subsection{Binary candidates}\label{sec.bincands}

The previous procedure yields 1\,181 positive flags that correspond to 1\,033 different stars since there are overlapping sources among the different criteria considered in this work. Of them, 43 are known binaries: 37 from {\it Gaia} variability analysis and 6 from the OGLE catalogues. The seventeen known binaries recovered in Simbad refer to one of those references. Unfortunately, these stars cannot serve as benchmarks for evaluating the remaining potential candidates for several reasons. On the one hand, the proposed criteria are not available for every star in the sample. For instance, not all binaries have GLIMPSE coverage to assess an excess in their SEDs nor have reduced chi-square values available: 19 of the 43 have a GLIMPSE counterpart and do not show IR excess, which can occur in equally bright binaries. Besides, among the 43 known binaries, only five fulfill one of the proposed criteria from \textit{Gaia}, being insufficient for a confident assessment. On the other hand, \textit{Gaia}'s astrometric parameters are not suitable for very tight pairs such as eclipsing binaries, as their short periods and close separations do not significantly affect the astrometric solution (e.g., \citealt{StassunTorres21,ElBadry24}). Alternatively, these astrometric disruptions may be related to a tertiary companion instead, capable of perturbing the motion of the inner binary and producing the observed anomalies \citep{Kostov25}.

Consequently, we kept all 990 stars (1\,033 - 43 known binaries) as potential binary candidates and identify them according to different lines of evidence:

\begin{enumerate}[label=\Roman*.]

\item Stars with IR excess in their SEDs constitute a well-constrained sample, as the excess provides a direct measurement of emission from a likely unresolved companion, since any background contamination has been rejected in Section~\ref{sec.IRexcess}.

\item Stars with reduced chi-square values (``$\chi^2$'') of 1 or 2 and those with variability flags are treated equivalently, as both indicate photometric variability. While such variability may be caused by a physical companion, it can also arise from intrinsic stellar phenomena such as magnetic activity, surface spots, or flares.

\item Stars with {\tt duplicated\_source} =1 (\ref{eq.dupsource}) that could be related to stellar multiplicity but also to difficulties on the processing of the astrometry or photometry.

\item Stars with {\tt ipd\_frac\_multi\_peak} > 30 (\ref{eq.multipeak}) and with {\tt ipd\_gof\_harmonic\_amplitude} > 0.1, {\tt astrometric\_excess\_noise} > 1.4, and {\tt astrometric\_excess\_noise\_sig} > 2 (\ref{eq.epsi}), which are linked to the presence of a visually resolved double star (e.g., \citealt{Tokovinin23,Holl23}).
\end{enumerate}

Moreover, within our binary sample we flag the strongest binary candidates by requiring them to present at least two \textit{True} flags out of those defined above. For this task, we combine the 47 stars with ``$\chi^2$'' variability flag with the seven stars with Gaia variability flag (``GVar'') and the six variables found in Simbad (``SVar''), since in all cases they rely on photometric variability. Then we select stars with two flags among IR excess, double peak (\ref{eq.multipeak}), duplicates source (\ref{eq.dupsource}), astrometric excess noise (\ref{eq.epsi}), and photometric variability (``$\chi^2$'' + ``GVar'' + ``SVar''). We ended up with 120 strong binary candidates. In addition, there are $5$ known binaries with astrometric (\ref{eq.dupsource} or \ref{eq.epsi}) flags that could indicate the presence of a tertiary companion. Except for these five, ``GBin'', ``OBin'' and ``SBin'' known binaries are not considered here and will only be used for the estimation of the multiplicity fraction. 

\section{Analysis and results}
\label{sec:analysisandresults}

\subsection{Contamination from background stars}\label{sec:contamination}

\begin{figure}
\centering
\includegraphics[width=0.45\textwidth]{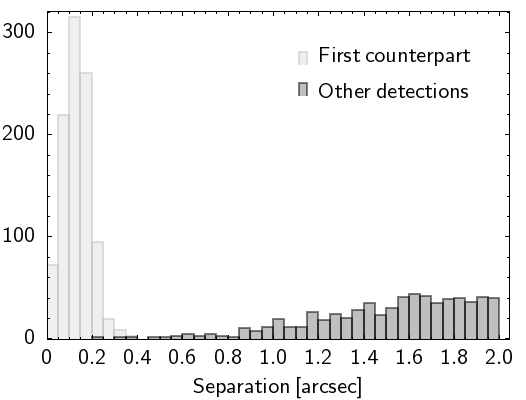}
\caption{Distribution of angular separations of the closest \textit{Gaia} counterpart for the binary candidates (grey) and other \textit{Gaia} detections within 2\,arcsec (black).}
\label{fig.histgaiaclosesep}
\end{figure}

\begin{figure}
\centering
\includegraphics[width=0.4\textwidth]{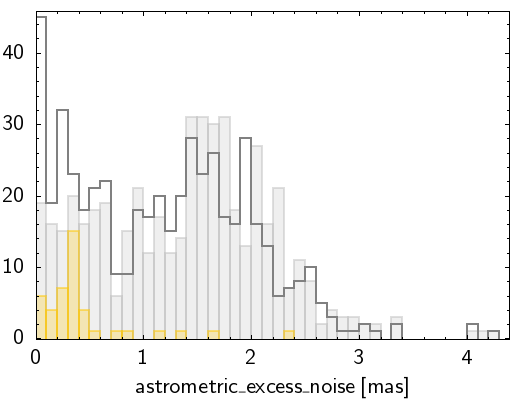}
\includegraphics[width=0.4\textwidth]{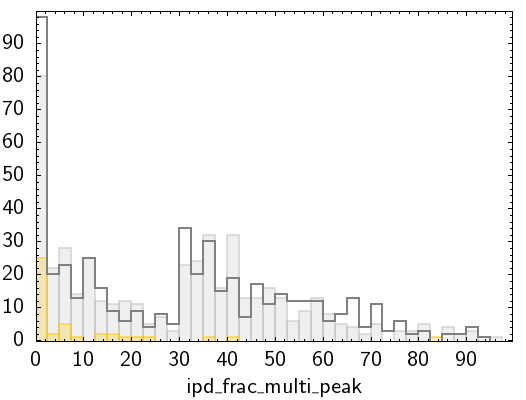}
\includegraphics[width=0.4\textwidth]{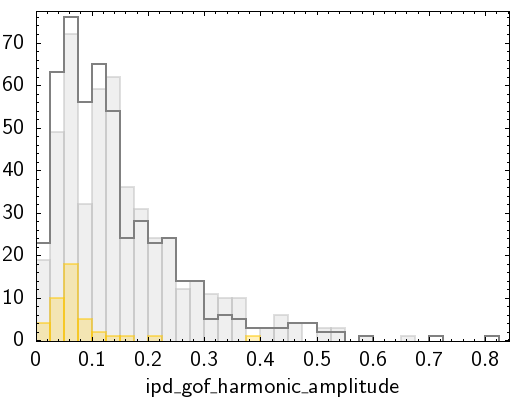}
\includegraphics[width=0.4\textwidth]{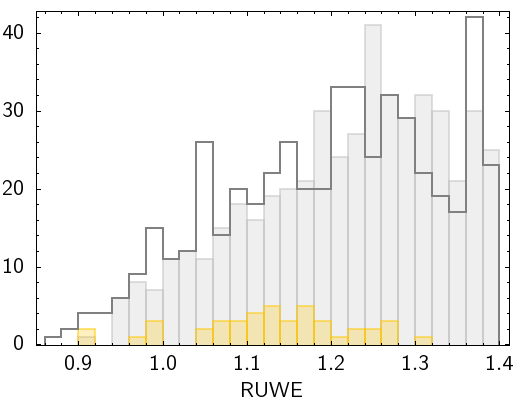}
\caption{Distributions of the {\tt astrometric\_excess\_noise}, {\tt ipd\_frac\_multi\_peak}, {\tt ipd\_gof\_harmonic\_amplitude} and {\tt RUWE} parameters. Solid line and filled bars stand for binary candidates with and without more than one \textit{Gaia} counterpart within 2\,arcsec, respectively. Yellow bars represent the sample of known binaries.}
\label{fig.histparams_closeneighbor}
\end{figure}

It is noticeable that, despite removing all stars with {\tt RUWE}>1.4 in PC, a number of known binaries and potential companions remain. Although the astrometric fit can be affected by the orbital motion of the pair, some authors have suggested that {\it Gaia} astrometric noise can also mask the wobble of the center of mass around the photocentre in a binary system, resulting in lower {\tt RUWE} values (e.g., \citealt{CastroGinard24}).
As for the {\tt ipd\_frac\_multi\_peak} parameter, it seems to be closely related to the presence of a close companion at less than 2\,arcsec (e.g., \citealt{Tokovinin23}). However, it does not evidence gravitational binding and can have high values for both, true binaries and chance alignments in high density environments \citep{ML23}.
In order to assess contamination due to the presence of a close star in the field that could affect the astrometric and photometric statistical parameters considered here, we first crossmatched the sample of candidates (including the strong candidates) obtained in the previous search with \textit{Gaia} DR3 and kept all counterparts within 2\,arcsec. Nearly half of the stars have more than one \textit{Gaia} detection in that radius, whose separations are shown in Figure~\ref{fig.histgaiaclosesep}. Our primary \textit{Gaia} counterpart lies within 0.4\,arcsec from our input coordinates taken from VVV. Very few of them have a visual neighbor within 1\,arcsec, including three that are closer than 0.37\,arcsec and that could be ascribed to a companion resolved with \textit{Gaia} (see Section~\ref{sec:properties}). In none of these cases does {\it Gaia} provide astrometric solutions that allow us to determine whether they share distance or proper motions.

Previous {\it Gaia} validation studies have shown that close neighbors can affect the astrometric solution at separations of a few arcseconds (e.g., \citealt{Lindegren21,Luna23}). Nearly half of our sample would be affected in this regime.
For assessing the influence of these visual companions on the {\tt astrometric\_excess\_noise} ($\epsilon$), the {\tt ipd\_frac\_multi\_peak} or even the {\tt ipd\_gof\_harmonic\_amplitude} and {\tt RUWE} parameters, we perform the non parametric Mann-Whitney U test \citep{Mann47} that applies for non-normal distributions. We define two samples from all the 990 (870+120) binary candidates with the 504 and 486 stars with and without other \textit{Gaia} detections within 2\,arcsec, respectively. 
We excluded the 43 known binaries from this analysis in order to avoid biases and for a better comparison between different indicators.
Although statistically significant differences (p-value<0.05) are found in {\tt astrometric\_excess\_noise} (p-value$=5.9 \cdot 10^{-5}$), {\tt ipd\_gof\_harmonic\_amplitude} (p-value=0.017903) and {\tt RUWE} (p-value=0.019722), the corresponding effect sizes are small, suggesting that the practical impact on the adopted selection criteria is negligible. As a matter of fact, all the three tend to be larger in the sample where there is no close neighbor, contrary to what would be expected. 
Our analysis is robust against contamination from field neighbors as close as 0.2\,arcsec or as bright as 15\,mag. The $G$-band magnitude contrasts with the target stars range from --3.4 to 3.9\,mag, with the brighter neighbors located at separations greater than 0.95\,arcsec.

Repeating the same analysis using only \textit{Gaia} counterparts within 1,arcsec, thereby restricting the search to the closest neighbors, results in a subset of 47 stars with more than one detection. No significant differences are found in any of the analysed parameters with respect to the remaining sample without additional \textit{Gaia} entries.
Our interpretation of these results is that, although there are statistically significant differences between candidates with and without {\it Gaia} neighbors at short angular separations (below 2\,arcsec), they do not have any impact on the parameters used and hence, on the cuts defined in this work for identifying binary candidates. When measuring such negligible impact, it points towards lower values of all the parameters when another star visually interferes, providing no evidence of contamination in these cases.
Figure~\ref{fig.histparams_closeneighbor} shows the distributions of the {\tt astrometric\_excess\_noise}, the {\tt ipd\_frac\_multi\_peak}, the {\tt ipd\_gof\_harmonic\_amplitude} and {\tt RUWE} parameters of the two samples defined. We also represent the set of known binaries for a direct comparison. They present similar trends than these two samples.

\subsection{Properties of the binary sample}\label{sec:properties}

\begin{figure}
\centering
\includegraphics[width=0.45\textwidth]{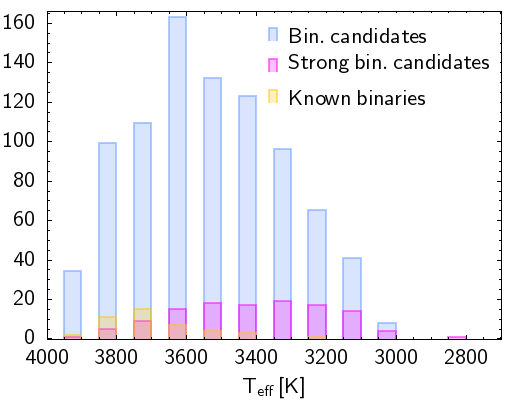}
\includegraphics[width=0.45\textwidth]{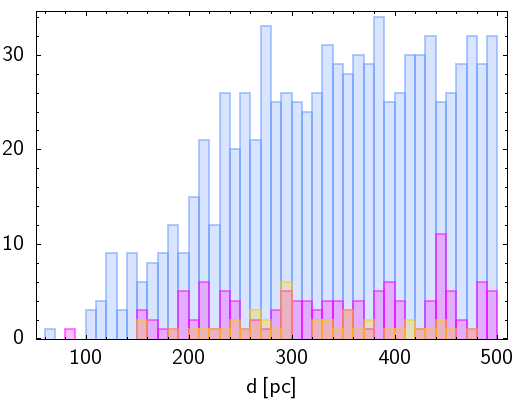}
\includegraphics[width=0.45\textwidth]{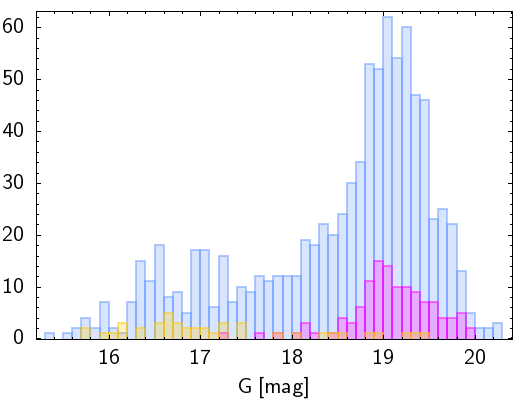}
\caption{Distribution of effective temperatures (top) and distances (middle) obtained from \citealt{PCruz23}, and \textit{Gaia}'s $G$ band magnitudes (bottom).}
\label{fig.histteffbins}
\end{figure}

\begin{figure}
\centering
\includegraphics[width=0.45\textwidth]{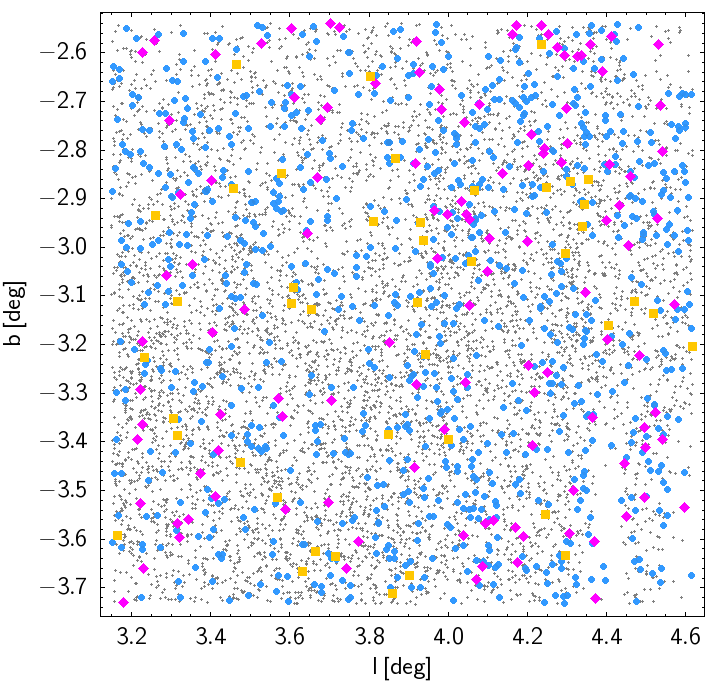}
\caption{Sky distribution of the 870 binary candidates (blue circles), the 120 strong candidates (magenta rhombs), and the 43 known binaries (orange squares), respectively. The 7\,925 M dwarfs of the sample are displayed with gray pluses.}
\label{fig.skydistbins}
\end{figure}

The 1\,033 candidates can be described in terms of their effective temperatures as derived with VOSA by \cite{PCruz23} as shown on the top panel in Figure~\ref{fig.histteffbins} separated by category: binary candidates, strong binary candidates, and known binaries (including here the triple candidates). Their distance and $G$-band magnitude distributions are presented on the middle and bottom panels of the same figure. The sample of binary candidates peaks at $\sim 3500$\,K (M2-M3 V), the strong candidates at $\sim 3400$\,K (M3 V) and known binaries at $\sim 3700$\,K (M1 V). Regarding distances, only two stars are closer than 100\,pc to the Sun: Gaia DR3 4063295969972933760 and Gaia DR3 4063027727739835008. Both of them are summarised in Table~\ref{tab.triples}.
The $G$-band magnitude distribution reveals that known binaries and strong binary candidates populate the brightest and the faintest ends of the distribution, respectively. This diagram discloses a certain bias towards faint stars, which have typically larger photometric errors and therefore larger values of the astrometric and photometric parameters compared to brighter stars as noticed by \cite{Lindegren21} and \cite{Riello21}. This likely has a greater impact on our candidate selection than the high stellar density of the field.
However, under these conditions we are unable to propose alternative constraints for the two criteria that contribute the largest fraction of candidates in our sample, namely criteria~\ref{eq.multipeak} and~\ref{eq.epsi}, without significantly compromising the completeness of the sample. Consequently, some level of contamination is expected to remain, and the nature of individual candidates can only be confirmed through dedicated follow-up observations.
We also observe that our binary sample has a higher average {\tt RUWE} value (1.22) than the remaining (presumably single) stars (1.08).
Their sky distribution is shown in Figure~\ref{fig.skydistbins}, where we do not observe any preferred direction like a star forming region. 

\begin{figure*}
\centering
\includegraphics[width=0.33\textwidth]{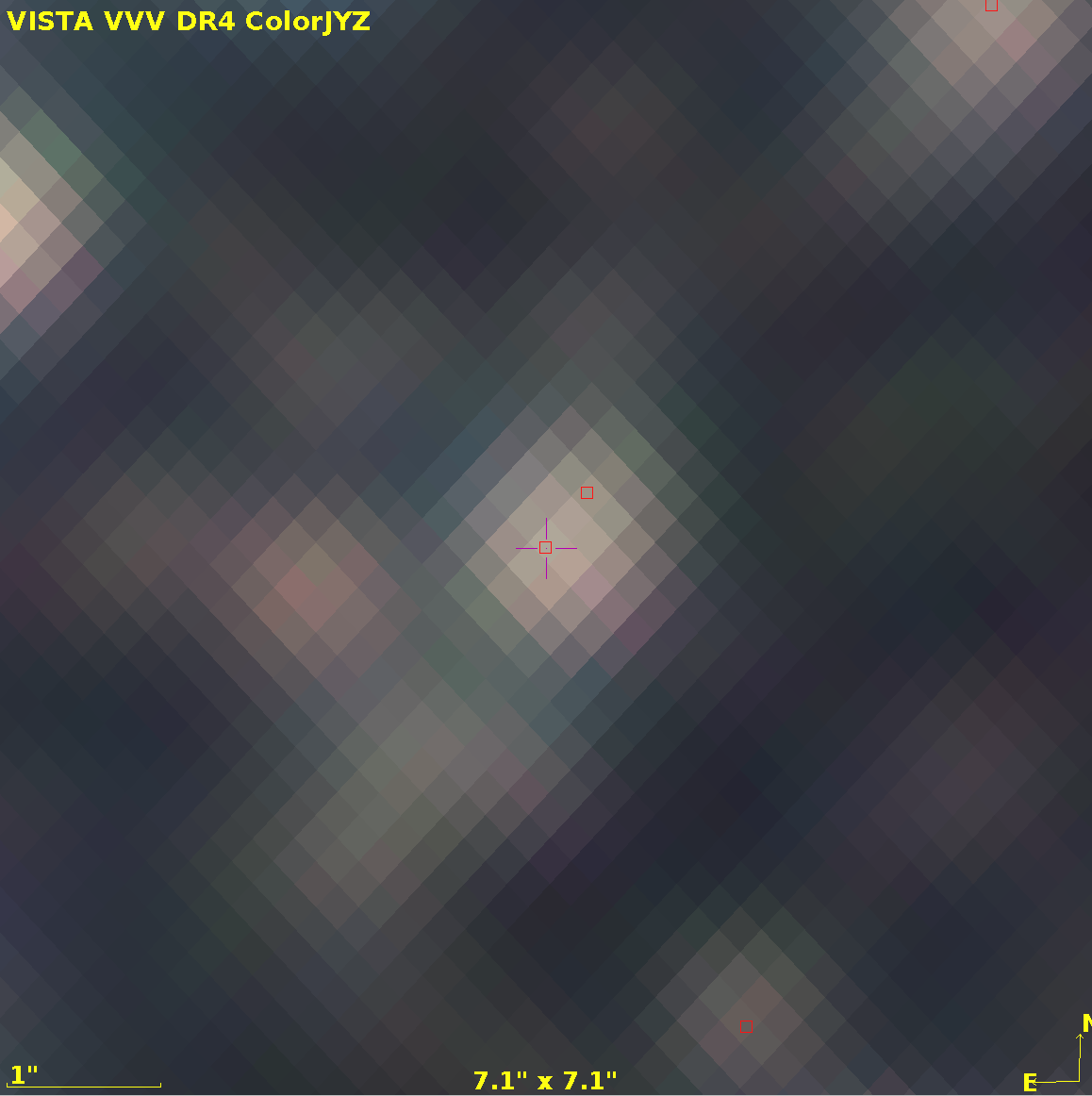}
\includegraphics[width=0.33\textwidth]{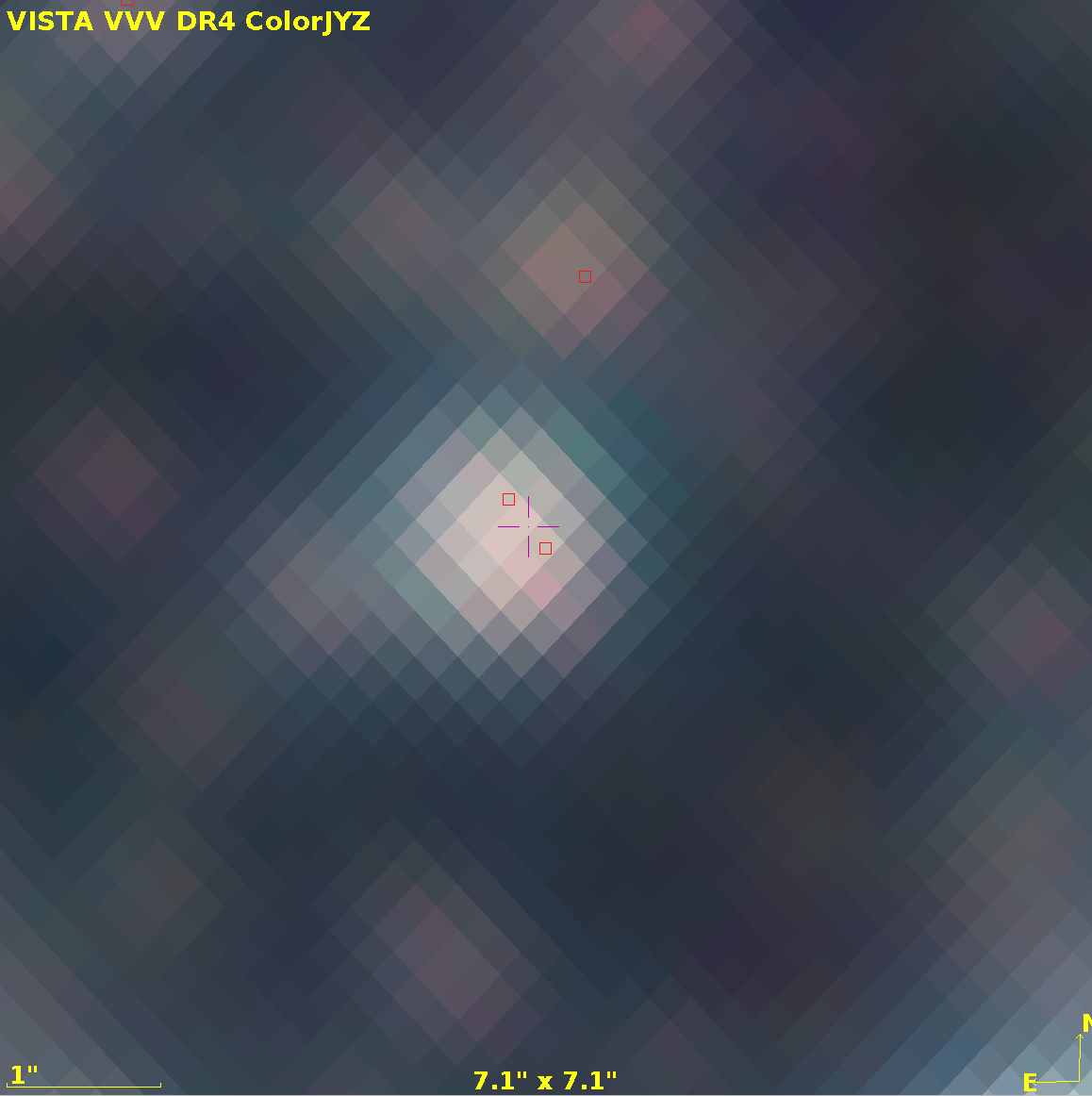}
\includegraphics[width=0.33\textwidth]{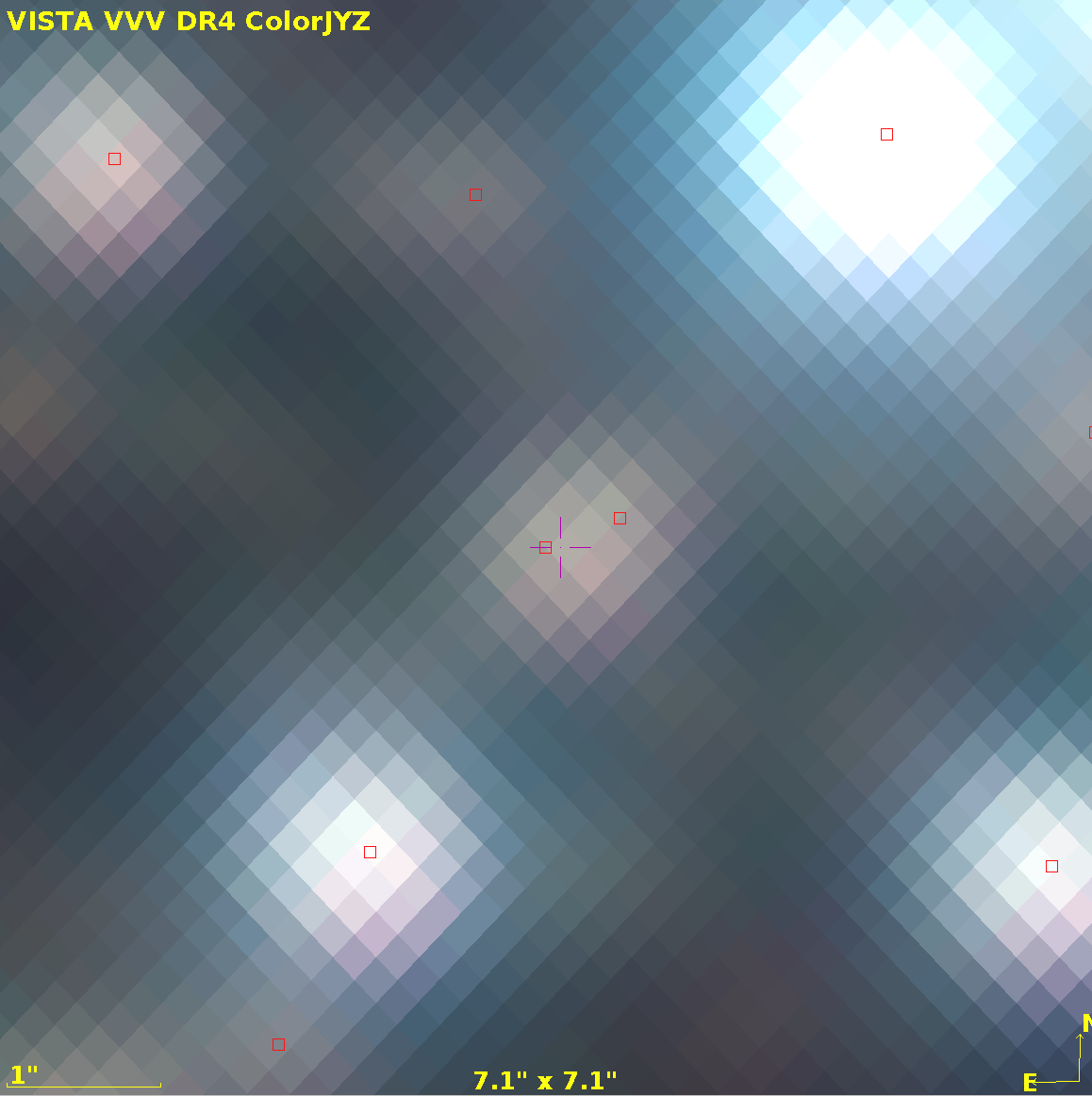}
\caption{Aladin sky view of the three potentially resolved companions: Gaia DR3 4063009066137457792, Gaia DR3 4063058101640048512, and Gaia DR3 4063295871084813568 from left to right. Magenta reticle shows the target position and the \textit{Gaia} detections are represented with red squares over the VVV DR4 colour image displayed in the background.}
\label{fig.aladin}
\end{figure*}

As noticed in Section~\ref{sec:contamination}, there are three binary candidates with a second \textit{Gaia} detection at less than 0.4\,arcsec that could be associated to a companion that is being resolved: Gaia DR3 4063009066137457792, Gaia DR3 4063058101640048512, and Gaia DR3 4063295871084813568. {\it Gaia} does not provide {\tt RUWE} or RVs for any of these secondary detections. Hence, we can only evaluate for them criteria~\ref{eq.multipeak}, \ref{eq.dupsource}, and \ref{eq.epsi} from Section~\ref{sec.gaiaindicators}. Only the secondary detection of Gaia DR3 4063058101640048512 (named Gaia DR3 4063058101699456512) fulfills criterion~\ref{eq.multipeak} related to the detection of multiple peaks in the IPD process. The other two detections, although presenting large {\tt ipd\_gof\_harmonic\_amplitude} or {\tt astrometric\_excess\_noise}, do not fulfill the combined selection criteria defined in Table~\ref{tab.Gaiacriteria}. They are included in Table~\ref{tab.triples} together with their angular separations and magnitude differences in the $G$ band, which range between 0.12 and 0.96\,mag. 

\textit{Gaia} reaches completeness when resolving stars separated by over 0.7\,arcsec but is able to resolve stars equally bright up to 0.4\,arcsec depending on the magnitude of the primary and the orientation along the dominating scan direction (see \citealt{Gaia21_Smart}). When using Eq.~2 of the cited work that provides an estimation of the minimum resolvable separation given the magnitude difference of the pair, we obtain minimum separations between 1.65 and 2.32\,arcsec, larger than the actual separations found for the detections around these three stars. This is because that equation applies for primaries as faint as 11\,mag, which is far too bright compared to our stars, for which \textit{Gaia} reaches even closer companions, as seems to be the case. Figure~\ref{fig.aladin} shows the three of them with a VVV image as a background and the two \textit{Gaia} detections.
For the remaining candidate systems in this work, we again visually inspected the VVV images and \textit{Gaia} detections using Aladin and found five other pairs of \textit{Gaia} detections separated by 0.6--0.7\,arcsec and magnitude differences between 0.1 and 1.2\,mag in the $G$ band. Bound companions at these separations and magnitude differences may not be responsible for the astrometric or photometric perturbations observed. Physical  binding of the detections could be confirmed via common proper motion or even orbital motion measurements. We do not find any other potential companion being resolved, meaning all of them are at much closer separations than 0.2--0.4\,arcsec, much fainter than our target star or both.

\begin{table*}
 \centering
 \caption{Candidate systems of interest.}
 \small
 \begin{tabular}{llcccccccc}
 \hline 
 \hline
 \noalign{\smallskip}
\textit{Gaia} DR3 Name  & Class$^a$ &  d [pc]$^b$  &   T$_{{\rm eff},1}$ [K]$^c$   &   T$_{{\rm eff},2}$ [K]$^c$   &  G3 &    G6   &   G8    & Separation [arcsec] &   $\Delta$G [mag]  \\

 \noalign{\smallskip}
 \hline
 \noalign{\smallskip} 

\multicolumn{9}{c}{\textit{Candidate binaries at less than 100\,pc}} \\
 \noalign{\smallskip}
  \hline
  \noalign{\smallskip}
4063295969972933760    &   Candidate   &   60.5 $\pm$ 6.0    &   3400    &...&...&$\mathsf{X}$&...&...&...\\
 \noalign{\smallskip}

4063027727739835008    &   Strong candidate    &   87.8 $\pm$ 5.4    &   3400    &...&   $\mathsf{X}$    &   $\mathsf{X}$    & ...&...&...  \\
 \noalign{\smallskip}
   \hline
  \noalign{\smallskip}
\multicolumn{9}{c}{\textit{Potentially resolved binary candidates with Gaia}} \\
 \noalign{\smallskip}
   \hline
  \noalign{\smallskip}

4063009066137457792 &   Candidate   &   273.9$^*$   &   3200    &...&    $\mathsf{X}$ &...& ...&   0.37$^d$    &   -0.96 $\pm$ 0.03\\
 \noalign{\smallskip}
4063058101640048512 &   Candidate   &   215.7$^*$   &   3100    &...&    $\mathsf{X}$ &...&...&0.23$^e$    &   -0.12 $\pm$ 0.02\\
 \noalign{\smallskip}
4063295871084813568    &   Candidate   &   463.0$^*$    &   3300    &...&    $\mathsf{X}$ &...&...&0.35$^f$    &   -0.53 $\pm$ 0.02\\
 \noalign{\smallskip}
  \hline
  \noalign{\smallskip}
  \multicolumn{9}{c}{\textit{Triple system candidates}} \\
   \noalign{\smallskip}
     \hline
  \noalign{\smallskip}
4062922002814215936   &   EB    &   359.4 $\pm$ 64.9 &    3700 &...&   $\mathsf{X}$   & ...&... &... &...\\
 \noalign{\smallskip} 
4063065416216940928     &   EB  &  438.0 $\pm$ 82.0    & 3800  &...& ...& ...  &   $\mathsf{X}$  &...&... \\
 \noalign{\smallskip} 
4063240135263100288     &   EB  &   359.1$^*$   & 3200  &...& $\mathsf{X}$   & ...&... &...&... \\
 \noalign{\smallskip} 
4063248622094959872     &ELL    &   187.5$^*$   & 3700  &...&  $\mathsf{X}$   & ... &... &...&...\\
 \noalign{\smallskip} 
4063315177052988544 &   EB  &  156.5 $\pm$ 20.5      &   3500 &...& ...& ...  &   $\mathsf{X}$  &...&... \\
 \noalign{\smallskip}
  \hline
  \noalign{\smallskip}
  \multicolumn{9}{c}{\textit{Strong binary candidates with IR excess}} \\
   \noalign{\smallskip}
     \hline
  \noalign{\smallskip}
4063128672309709440 & Strong candidate  &  451.2$^*$   &   3500$^*$    &   1200$^*$    &   $\mathsf{X}$    & ...  & ...  &...&...\\
\noalign{\smallskip}
4063155923861864448 & Strong candidate  &   489.1$^*$   &   3900$^*$    &   3500$^*$    &  ...  &...   & $\mathsf{X}$   &...&...\\
\noalign{\smallskip}
4063160867493208064& Strong candidate  &  384.0$^*$   &   3600$^*$    &   3100$^*$    &   $\mathsf{X}$    & ...  &...   &...&...\\
\noalign{\smallskip}

4063241788889479680& Strong candidate  &  347.7$^*$   &   4000$^*$    &   3600$^*$    &   $\mathsf{X}$    & ...  &...   &...&...\\
\noalign{\smallskip}
4063246289960753024& Strong candidate  &  458.6$^*$   &   3600$^*$    &   3400$^*$    &   $\mathsf{X}$    &  ... &...   &...&...\\
\noalign{\smallskip}

4063285180944001792& Strong candidate  &  401.2$^*$   &   3400$^*$    &   1300$^*$    &   $\mathsf{X}$    & ...  &...   &...&...\\
\noalign{\smallskip}
4063340156572112768& Strong candidate  &   305.7$^*$   &   3200$^*$    &   3100$^*$    &  ...  & ...  & $\mathsf{X}$   &...&...\\
\noalign{\smallskip}
4063344215192994816& Strong candidate  &  290.4   &   3300$^*$    &   3100$^*$    &   $\mathsf{X}$    &  ... & ...  &...&...\\
\noalign{\smallskip}
4063346994141650944& Strong candidate  &   488.0   &   3400$^*$    &   1200$^*$    &  ...  &  ... & $\mathsf{X}$   &...&...\\
\noalign{\smallskip}
 \hline
 \end{tabular}
\vspace{2mm}
\begin{minipage}{\textwidth}
{\footnotesize
$^{a}$ Classifications of known systems from \textit{Gaia}. EB: Eclipsing Binary; ELL: Ellipsoidal variable.
$^{b}$ Distances with ``*'' are spectophotometric and from \textit{Gaia} otherwise (see \citealt{PCruz23}).
$^{c}$ Effective temperatures with ``*'' are obtained from the binary fit with VOSA and taken from \cite{PCruz23} otherwise. Error-bars are half of the used grid of models in both cases: 50\,K.
$^{d}$ Resolved object is Gaia DR3 4063009070278273792.
$^{e}$ Resolved object is Gaia DR3 4063058101699456512.
$^{f}$ Resolved object is Gaia DR3 4063295871084813440.
}
\end{minipage}
\label{tab.triples}
\end{table*}

\begin{figure}
\centering
\includegraphics[width=0.45\textwidth]{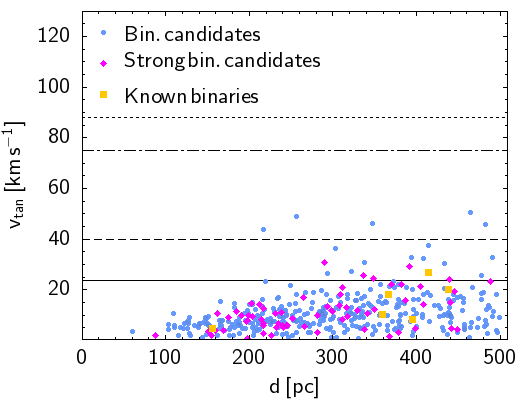}
\caption{Tangential velocity vs. distance. Color code is the same as in previous figures. Solid, dashed, dot-dashed and dotted black lines represent the mean values of $v_{\rm tan}$ of the young, thin, thick-thin, and thick discs populations from \citealt{CC24}.}
\label{fig.vtan}
\end{figure}

From the PC catalogue, we select among the 1\,033 binaries and binary candidates the 422 objects that have distances and proper motions gathered from \textit{Gaia} and with good astrometric solutions as defined in \citealt{PCruz23}.
With them, we compute their tangential velocities $v_{\rm tan} [{\rm km}\,{\rm s}^{-1}] = 4.74 \cdot d\,[{\rm pc}] \cdot \mu\,[{\rm arcsec\,a^{-1}}]$ and evaluate their distribution in the different stellar Galactic populations (thin and thick discs, transition between them, and halo), including the young disc population as defined by \cite{Leg92}. For that, we used the median reference values for each population from \cite{CC24}. In Figure~\ref{fig.vtan}, we displayed the 422 stars colour coded according to their classifications as known binaries (6), and binary/strong binary candidates (350/66). Almost all of them (95\%) have low tangential velocities compatible with the young disc population ($\tau \lesssim 1$\,Ga), where it also lies the \textit{Gaia}'s YSO candidate Gaia DR3 4063086856413982080. For this, we used 24\,km\,s$^{-1}$ as an upper limit. This is in fair agreement with the region of interest, close to the Galactic plane. The position of the remaining stars is coherent with a thin disc population, having tangential velocities up to 50\,km\,s$^{-1}$.

In an attempt to complement this analysis, we use the {\tt BANYAN}~$\Sigma$ algorithm \citep{Gagne18} that can predict membership to 32 stellar associations within 150\,pc from the Sun or to the Galactic field neighborhood within 300\,pc even lacking radial velocities. The code returns high probabilities of field membership for all binary candidates within these limits. No further kinematic analysis can be performed because \textit{Gaia} does not provide radial velocities for these stars.

\subsection{Known binaries}\label{sec:known}

In our sample there are 43 binaries identified either with \textit{Gaia} or with the OGLE project: 25 ellipsoidal variables, 17 eclipsing binaries and one RS CVn type. 
The six EBs with computed tangential velocities are compatible with a young disc population (see Figure~\ref{fig.vtan}).

In Figure~\ref{fig.histparams_closeneighbor} we observe that the sample of 43 known binaries typically lie at the lower end of the histograms except for {\tt RUWE}, were they spread and present values mostly over 1. This means that a large fraction of them would remain unidentified based on our criteria using these statistical parameters and that there could be a non negligible number of true binaries hidden under low {\tt ipd\_frac\_multi\_peak}, {\tt astrometric\_excess\_noise} or {\tt ipd\_gof\_harmonic\_amplitude} values. Nonetheless each of these parameters can still provide signs of the presence of an unresolved wider companion, turning the system into a triple: there are $3+2$ binaries with {\tt ipd\_frac\_multi\_peak} > 30 and {\tt astrometric\_excess\_noise} > 1.4, respectively. They are listed in Table~\ref{tab.triples}.

\subsection{Candidates with IR excess}

Of the total of 990 binary candidates (excluding the 43 known binaries from the sample of 1\,033 stars described in Sect.~\ref{sec.bincands}), there are 98 stars with an IR excess in their photometric SEDs obtained with VOSA in \cite{PCruz23}. For them, we perform a binary model fit of their SEDs using VOSA, aiming at characterizing the secondary candidate companion. Similarly as in \cite{PCruz23}, we used the ZYJHKs photometry from VVV, and searched for \textit{Gaia} DR3, VPHAS+DR2 \citep{Drew14}, and GLIMPSE Source Catalog I + II + 3D photometry within 1\,arsec using VOSA. Extinction was set to range between 0 and 0.5\,mag, surface gravity ($\log{g}$) between 4 and 5.5\,dex and the chosen model was BT-Settl CIFIST, constrained to solar metallicity [Fe/H]=0.

\begin{figure}
\centering
\includegraphics[width=0.55\textwidth]{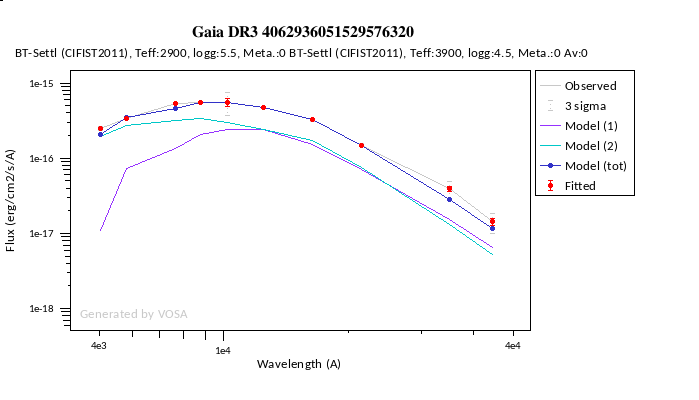}
\includegraphics[width=0.45\textwidth]{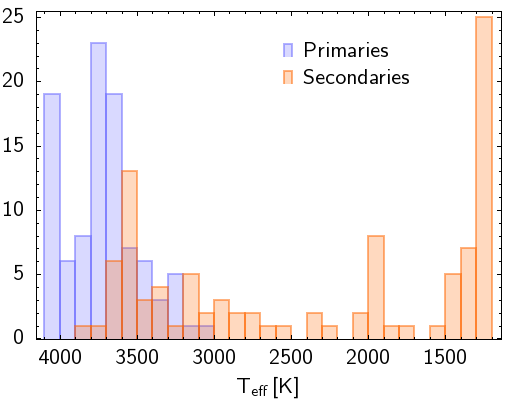}
\caption{\textit{Top:} Example of SED binary fitting using VOSA. Blue and red filled circles represent the synthetic photometric calculated from the best-fit model and the observed photometry, respectively. Grey and blue solid lines show the stellar observed and modeled SED of the composed system, while cyan and purple solid lines represent the modeled stellar SED of the primary and secondary components, respectively.}
\label{fig.vosabin}
\end{figure}

VOSA provides a modified reduced $\chi^2$ statistic, denoted Vgf$_b$, in which small observational uncertainties are replaced by a value equal to 10\% of the observed flux. This parameter is intended as a visual goodness-of-fit estimator that reduces the influence of underestimated photometric uncertainties, although the best-fitting model is still selected using the standard reduced $\chi^2$\footnote{See the VOSA documentation: \url{http://svo2.cab.inta-csic.es/theory/vosa/helpw4.php?otype=star&action=help&what=fit}}. Values of Vgf$_b \lesssim$ 12--15 generally indicate a satisfactory fit to the observed SED.
We obtained fits with Vgf$_b$<15 for all but three binary systems, which have slightly higher Vgf$_b$ values up to 24.
We keep all of them for this study. An example of a binary SED fitting is shown on top of Fig.\ref{fig.vosabin}, together with the effective temperature distributions of the primary and secondary components. While primaries range from 3000 to 4000\,K, secondaries spread from 3700 to 1200\,K reaching down into the L and T regimes. The adopted errors are half of the grid models: 50\,K. It is worth emphasizing that the obtained $T_{\rm eff}$ are initial estimates and they should be further confirmed with spectroscopic methods.

We can tentatively separate the systems into those composed by two M dwarfs (M+M), an M dwarf plus an L dwarf (M+L), and an M dwarf plus a T dwarf (M+T) using as a reference the spectral type--effective temperature relation from the public updated table of \cite{Pecaut13}\footnote{\url{https://www.pas.rochester.edu/~emamajek/EEM_dwarf_UBVIJHK_colors_Teff.txt}}. According to this, there are 47, 19 and 32 M+M, M+L and M+T systems, respectively. This implies that at least one third lie at the substellar regime. The primaries are all earlier than M5\,V.
Interestingly, there are nine of these systems flagged as strong binary candidates that are included in Table~\ref{tab.triples}: six M+M and three M+T systems. 

\subsection{Multiplicity fraction}

Observationally, the multiplicity fraction in M dwarfs ranges between 20 and 32\% (e.g., \citealt{Bergfors10,Janson12,Jodar13,Winters19,GP26}), although several authors suggest that it may reach up to $\sim$40\% when unresolved candidate binaries are included (e.g., \citealt{Cifuentes25} and references therein). Assuming all our candidates are true binaries, we derive a binary fraction of 13\% in the PC sample, whereas known binaries and strong binary candidates alone represent only 2\% of the sample.

This estimate reflects, among others, the fact that the PC sample is intrinsically biased as it intentionally excluded stars with RUWE>1.4. Moreover, our methodology is restricted to unresolved systems that are sufficiently wide to produce detectable astrometric or photometric perturbations while still allowing the identification of a composite spectral energy distribution. It therefore misses the tightest binaries such as EBs, which may remain undetected when the statistical parameters used in this work exhibit low values (see Figure~\ref{fig.histparams_closeneighbor} and Section~\ref{sec:known}), and does not account for resolved nor wide binaries. We must also take into account the faintness of our targets and their companions, whose masses can be below the hydrogen-burning limit. This makes it difficult to detect perturbations in their motion across the sky, even with the highly precise astrometry from \textit{Gaia}, which is optimized for solar-type and more massive stars that are bright in the optical. Some of these systems could instead be resolved through adaptive optics or lucky-imaging observations.

In addition, the binary assessment is incomplete because not all stars in the sample have the ancillary information required by our methodology. As already commented in Section~\ref{subsecIR}, not every object has VVV light curves, GLIMPSE coverage for assessing an IR excess. Also not all the OGLE catalogues consulted spread over the entire region where the b294 tile lies. Consequently, nearly half of the PC sample cannot be fully evaluated using all the binary diagnostics considered in this work.
To minimise this incompleteness, we restrict the analysis to the region covered by GLIMPSE and the OGLE binary catalogue of \citet{Soszynski16}. This leaves a homogeneous sample of 4065 M dwarfs, all with evaluable \textit{Gaia} indicators, from which we identify 606 binary candidates and known binaries (67 strong candidates and 27 known binaries). This corresponds to a binary fraction between 2.3 and 14.9\%, only slightly higher than that obtained for the full sample.

For comparison, \citet{Winters19} measured a close multiplicity fraction of 20.2\% for M dwarfs at  projected separations below 50\,au. This limit was adopted because of the non-uniform sensitivity of their survey at separations below 2\,arcsec. As discussed before, \textit{Gaia} is able to resolve companions down to angular separations of $\sim$0.4\,arcsec. At the maximum distance of our sample (500\,pc), a projected separation of 50\,au corresponds to only 0.1\,arcsec, implying that such systems would remain unresolved by \textit{Gaia} and therefore lie within the separation regime targeted by our methodology. Although the two studies are not directly comparable owing to their different selection criteria, distance ranges, and observing environments, the difference between the close multiplicity fraction reported by \citet{Winters19} (20.2\%) and our highest recovered fraction (14.9\%) suggests that our methodology may miss, at least, of the order of one quarter ($\sim$26\%) of close binary systems.

Several factors probably contribute to this remaining discrepancy.
First, the presence of a close companion does not always produce measurable perturbations in the \textit{Gaia} astrometric and photometric parameters beyond RUWE. This impact depends primarily on the angular separation and relative brightness of the companion, and secondarily on other factors such as the orbital properties of the system (e.g., eccentricity and orbital-plane orientation). Any measurable influence becomes increasingly difficult to discern in crowded regions. Second, the infrared-excess method is itself incomplete and is expected to miss a non-negligible fraction of binary systems. In M dwarf binaries, the companion mass-ratio distribution is skewed toward high values (q$\gtrsim$0.7, \citealt{DK13,Winters19}), implying that both components frequently have similar effective temperatures and spectral energy distributions. Consequently, the additional flux contributed by the secondary produces little or no detectable infrared excess, making such systems difficult to identify through SED fitting alone.

Our results are nevertheless consistent with those of \cite{Cifuentes25}, who obtained an unresolved binary fraction of 12.6\% (278/2214) for nearby M dwarfs using criteria ~\ref{eq.ruwe} to \ref{eq.nonsingle}. Pending confirmation of the candidates, we therefore interpret the recovered fraction of $\sim$15\% as the maximum close binary fraction recoverable with this methodology.

Finally, it should be noted that the surveys aimed at obtaining complete multiplicity censuses are typically limited to distances of 10, 25, or at most 50\,pc, whereas our sample extends to 500\,pc. Since none of our binary candidates lie within these nearby volumes, a direct comparison of multiplicity distributions is not possible. Instead, our catalogue complements previous multiplicity surveys by providing a sample of candidate close binaries at substantially larger distances. 
The absence of pairs at closer distances is explained by the low number of M dwarfs within 100\,pc in the PC catalogue. Although the b294 tile samples only a small region of the Galactic plane and bulge, this deficit may hint at variations in the stellar density of low-mass stars toward these regions. In the context of this work, however, the effect is more likely due to a combination of survey footprint, observational biases, and catalogue selection effects rather than a true astrophysical absence, although this should be further investigated in light of previous studies (e.g., \citealt{YannyGardner13,Ferguson17,Warren21}).

\section{Summary and conclusions}
\label{sec:summary}

We used the catalogue of M dwarfs within 500\,pc built with VVV photometry in \cite{PCruz23} to search for close (unresolved) companions that survived the {\tt RUWE}< 1.4 cut using public data. We evaluated several \textit{Gaia} statistical parameters commonly used to identify potential binary systems based on their astrometry and photometry. We also investigated the IR excess observed in the spectral energy distributions of some stars in the catalogue, previously identified in \cite{PCruz23}, and complemented the study with a literature search.

We obtained a sample of 1\,033 M dwarfs with known and suggested binary (or tertiary) companions, classified according to different aspects such as the evidence of emission of a cooler body, the fulfilling of astrometric criteria or variability:

\begin{itemize}

    \item 42 are known ellipsoidal or eclipsing binaries, and one is a RS CVn type binary. Five of them are proposed to be triple systems.
    
    \item 990 are binary candidates that satisfy at least one of the criteria defined for binarity assessment. Three of them are likely being resolved with \textit{Gaia} and 120 are presented as the best candidates that fulfill more than one criterion.
\end{itemize}

They spread from M0 to M5\,V and are beyond 100\,pc except for two binary candidates.
We were able to characterize reliable binary companion candidates of 98 stars with IR excess using VOSA, refined after visual inspection: all primaries are earlier than M5\,V (T$_{\rm eff}$>3000\,K) and secondaries reach the L and T dwarf regime with effective temperatures down to 1200\,K. We tentatively estimated that there are 47, 19 and 32 M+M, M+L and M+T systems, respectively. We highlight the nine systems among them that are flagged as best candidates.

Almost all of the 422 confirmed and candidate binaries, for which we could compute tangential velocities are kinematically consistent with a young disc population ($\tau \lesssim 1$\,Ga).

We evaluated the impact that close visual neighbors within 2\,arcsec detected with \textit{Gaia} have on the statistical parameters used in this work and conclude that the influence is negligible. Hence, the application of lower limits regarding them has been proven to be still valid in dense fields such as the b294 tile.

Finally, we obtained a multiplicity fraction of unresolved systems between 2 and 13\% that increases to 2.3 and 14.9\% when limiting the sample to the region covered by all the available studies considered. Although the identification of binary systems is restricted to publicly available data and the parameters used as binary tracers are blind to the shortest separated systems as eclipsing binaries and systems separated wide enough to prevent the excess detection in the SED, this is the first observational estimation of the fraction of unresolved systems in the M dwarf regime towards the galactic bulge.

The methodology, based on astrometric and photometric criteria and previously shown to be reliable for field stars, has been applied to denser stellar regions, albeit with stricter constraints. It reveals the presence of close binaries hidden among sources with low {\tt RUWE} values and identifies a sample of binary candidates for follow-up. If confirmed, these systems will further validate the methodology and help quantify the false-positive rate. Applied to such regions, the method yields promising results and highlights a region of interest that remains largely unexplored.

We emphasize the large distances covered in this work compared to other distance-complete studies in the solar vicinity. For a proper comparison, further research is needed in this direction, extending to greater distances and exploring different Galactic latitudes. This approach will also help to probe potential variations in stellar densities along different lines of sight from the Sun. Such studies will require challenging high-resolution imaging and spectroscopic observations in dense stellar fields, not yet accessible with current instrumentation. The forthcoming Gaia DR4 release, including more non-single-star solutions, astrometric and photometric epoch data, and radial velocities for an even larger number of stars, will provide increased leverage for the stellar characterization of vast and densely populated regions accessible for the community. In this regard, particular interest is focused on NIR parallax programs, such as the ground-based VLT/HAWK-I and GTC/EMIR, as well as the space-based HST/WFC3, JWST/NIRCam, and the proposed GaiaNIR mission, which are expected to enhance our understanding of the low-mass end of the HR diagram and expand the census of binary and multiple systems, thereby improving constraints on their formation and fundamental parameters.

\section*{Acknowledgements}

We acknowledge financial support from the Agencia Estatal de Investigaci\'on (AEI/10.13039/501100011033) of the Ministerio de Ciencia e Innovaci\'on and the ERDF ``A way of making Europe'' through projects
PID2022-137241NB-C4[4].
This work was supported by the Convenio between the Comunidad de Madrid and the Universidad Complutense de Madrid through the public call for grants for projects led by emerging researchers (Project No. PR17/24-31903, ref. 4331903). This research has made use of the Spanish Virtual Observatory (https://svo.cab.inta-csic.es) project funded by MCIN/AEI/10.13039/501100011033 through grant PID2023-146210NB-I00.

This publication makes use of VOSA, developed under the Spanish Virtual Observatory (https://svo.cab.inta-csic.es) project through grant PID2023-146210NB-I00 funded by MICIU/AEI/10.13039/501100011033 and by ERDF/EU. 
This research made use of the CDS cross-match service \citep{CDSBoch12,CDSPineau2020}, SIMBAD database \citep{Wenger00}, VizieR catalogue access tool \citep{Och00}, and Aladin sky atlas \citep{Bonnarel00, BF14} provided by CDS, Strasbourg, France. This research also made use of the TOPCAT \citep{Taylor05}.
The authors used ChatGPT (OpenAI) to assist with English language editing and manuscript readability. All scientific content, analyses, interpretations, and conclusions were developed and verified by the authors.

\section*{Data Availability}

All results obtained in this work are available for the benefit of the astronomical community.
Tables can be accessed from a webpage\footnote{\url{http://svocats.cab.inta-csic.es/mdwarfbin_vvv}} or through a Virtual Observatory ConeSearch\footnote{\url{http://svocats.cab.inta-csic.es/mdwarfbin_vvv/cs.php?RA=271.877&DEC=-28.079&SR=0.1&VERB=2}}, as well as from VizieR.
The archive provides a straightforward search interface, enabling queries by coordinates and radius, as well as by other parameters of interest. Query results are presented as an HTML table listing all sources that match the search criteria, and can also be downloaded in VOTable or CSV format. The archive also supports the SAMP (Simple Application Messaging Protocol) standard, allowing Virtual Observatory applications to communicate seamlessly and transparently with one another.
Tables will also be available in electronic form at the CDS.

We aim at contributing to increase our knowledge on the field with ready to use open data and a proven methodology based on the use of Virtual Observatory tools and services.



\bibliographystyle{mnras}
\bibliography{biblio} 



\appendix

\section{Pro-Am collaboration}\label{app.proam}

A Pro-Am collaboration is the result of joint efforts between professional and amateur astronomers to develop a science case. These collaborations reflect the enthusiasm of non-professional astronomers and the interest of professionals in opening science to them. They are part of national astronomical societies that perform under the umbrella of the Federaci\'on de Asociaciones Astron\'omicas de España (FAAE\footnote{\url{https://www.federacionastronomica.es/}}).
Equivalently for the professional community, the Sociedad Española de Astronom\'ia (SEA\footnote{\url{https://www.sea-astronomia.es/}}) promotes the development of astronomy and astrophysics in Spain. It hosts an active Pro-Am commission\footnote{\url{https://proam.sea-astronomia.es/}} devoted to serve as a link between the professional and amateur communities. It provides resources and tools for both, and showcases the results generated by these fruitful collaborations. Among other activities, the Pro-Am Commission organizes regular live-streamed public talks to disseminate past, ongoing, and future Pro-Am collaborations. These are available on the FAAE’s YouTube channel\footnote{\url{https://www.youtube.com/c/FAAE-Astronom\'ia/streams}}. On December 2024, we presented our project in this context, having more than 380 visualizations at the date of this publication. From that day, we formed a working group that summed 24 amateur members willing to contribute to the project.

\begin{figure}
\centering
\includegraphics[width=0.3\textwidth]{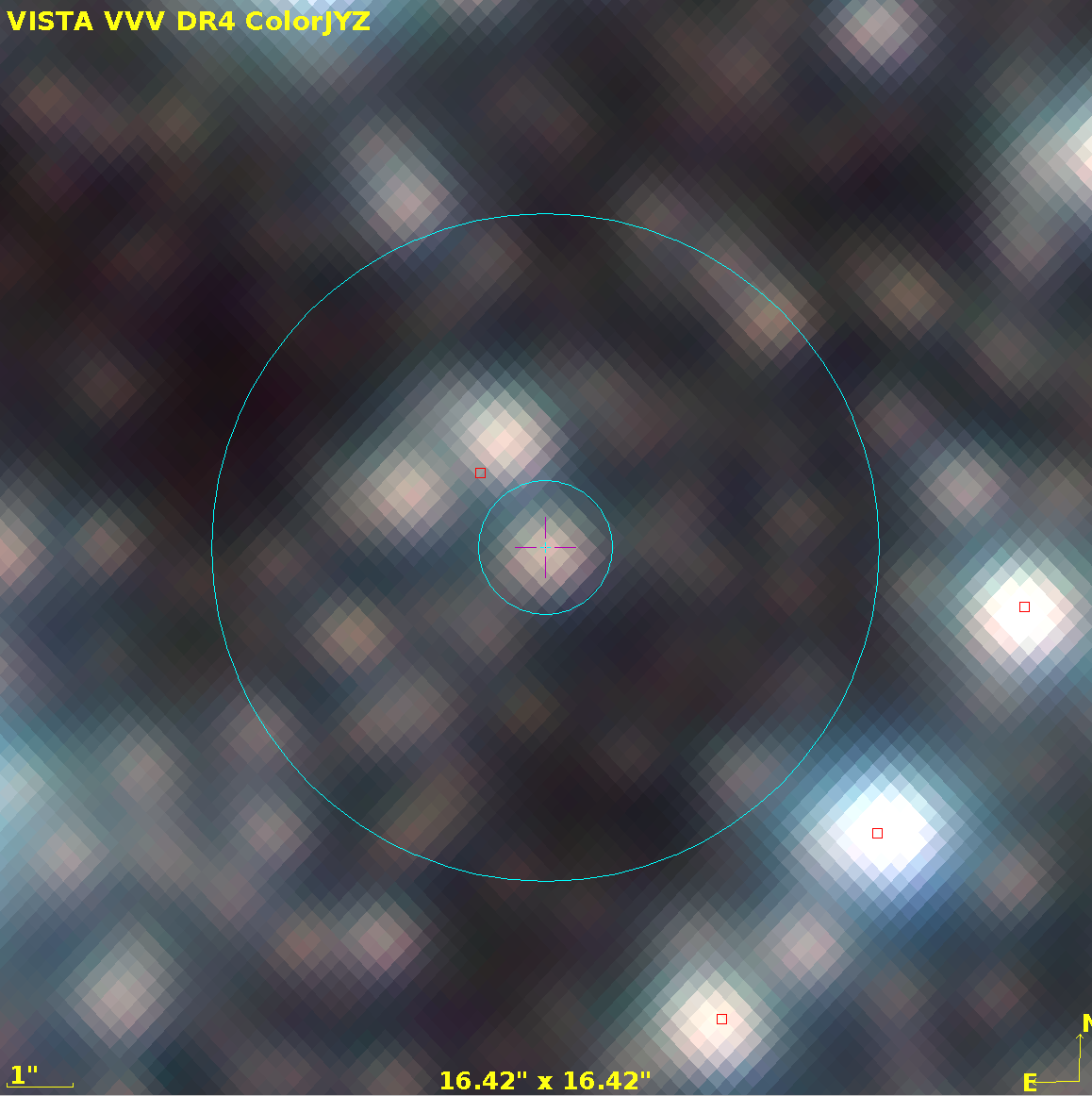}
\includegraphics[width=0.3\textwidth]{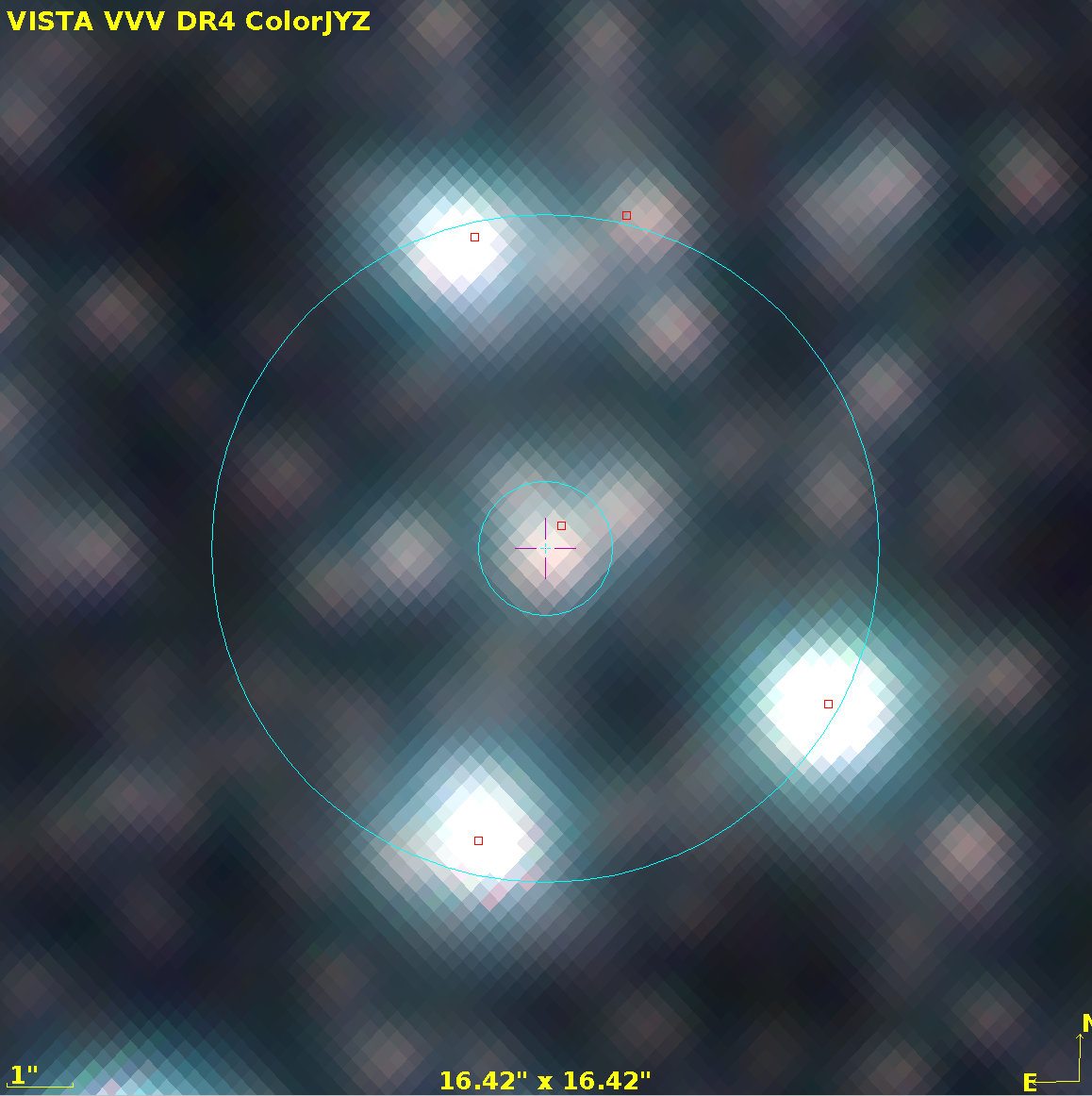}
\includegraphics[width=0.3\textwidth]{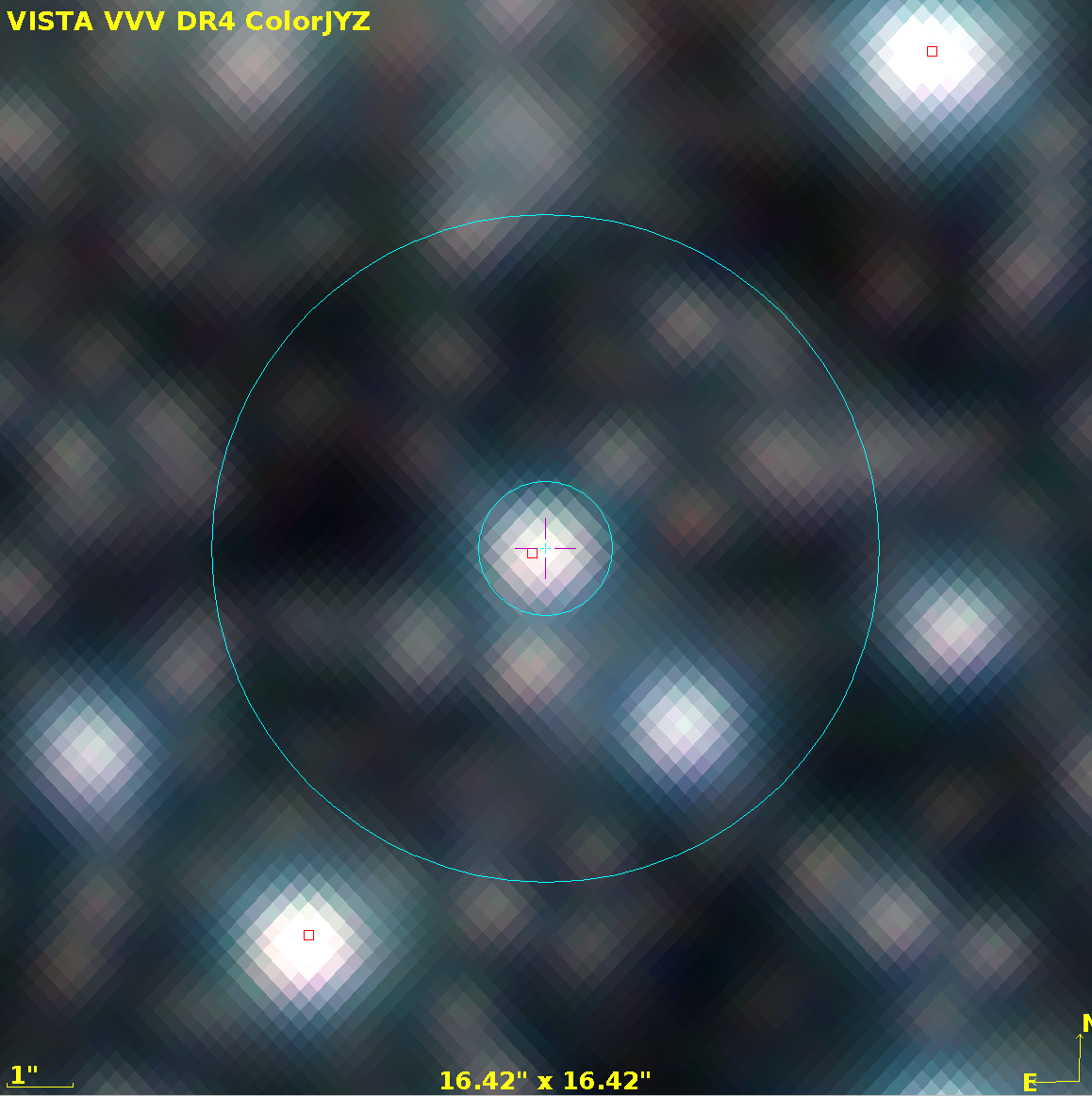}
\caption{Examples of stars among the 1\,228 sources with IR excess in their SEDs. Up, middle and bottom show a contaminated, doubtful and valid GLIMPSE counterpart, respectively. Magenta crosses represent the input coordinates of VVV, red open squares are the GLIMPSE photometric counterpart and the small and large cyan circles have radii of 1 and 5 arcseconds, respectively. The VVV DR4 colour image is displayed in the background.}
\label{fig.ejemplos}
\end{figure}

As already noted, crowded fields represent a challenge when using photometry for stellar characterization. The presence of close neighbors can contaminate the photometry of our target star, yielding wrong conclusions. The proposed science case uses a sample of 1228 M dwarfs with IR excess in their SEDs. The main objective of this exercise was to determine whether this excess was truly related to the target star or came from a nearby redder background star. In order to clean our sample from these background contaminants, we developed a methodology based on the visual inspection of the field using the Aladin Sky Atlas \citep{Bonnarel00}. It allows us to easily perform a repetitive task though a script that takes as an input the star coordinates to be displayed in a VVV DR4 colour image and two concentric circles with radii 1 and 5\,arcsec around the star's coordinates. In addition, it loads the GLIMPSE catalogue in the region in order to visually confirm that the GLIMPSE detection ascribed to the IR excess belongs to the target star or, on the contrary, reject it. To do so, we defined three different classes: valid GLIMPSE counterparts (``Y''), contaminated (``C''), and doubtful (``D''). Examples of these three cases are shown in Figure~\ref{fig.ejemplos}, where we represent the VVV DR4 colour image, the position of the target star and the positions of the GLIMPSE detections in a field of 16$\times$16\,arcsec$^2$. Typically, a GLIMPSE counterpart outside the 1\,arcsec radius circle is related to a field star, being classified as a contaminated source of excess (top panel). On a different scenario, where the GLIMPSE counterpart lies within the 1\,arcsec radius circle, the interpretation is not simple, since other nearby background stars could be contributing as well to the IRAC photometry (middle panel).
Disentangling the degree of photometric contamination is not straightforward and depends on both the offset between the GLIMPSE source and the centre of the M dwarf, and the proximity of neighbouring sources in the VVV image whose flux falls within the 1\,arcsec circle. Objects for which contamination cannot be ruled out, but can not be established unambiguously, are flagged as doubtful. The bottom panel illustrates an example of a valid GLIMPSE counterpart.

Given that this classification relies solely on the apparent brightness of the stars and their angular separation, and due to its by-eye nature, each star was inspected by five to six independent observers to reach overall agreement on the classification.
Each of them filled in a formatted table with their classifications (Y, C, D).
In total, we gathered 5\,750 classifications for the 1228 stars. Only stars with consensus over 60\% were classified as ``C'' (contaminated) or ``Y'' (valid) excess. Stars without sufficient consensus and those with doubtful assignment were flagged as inconclusive (``I''), yielding to three different classes used for carrying out the science case presented in this work. The justification of this careful exercise is to avoid misclassifications. The overall collaboration has yielded fruitful discussions and provided valuable results which serve for the purposes of this work.

\bsp	
\label{lastpage}
\end{document}